\documentclass[prd,aps,a4paper,eqsecnum,floatfix,nofootinbib,twocolumn]{revtex4}  %

\usepackage{graphicx} 
\usepackage{amsmath,amsfonts,amssymb}
\usepackage{tikz}
\usepackage{appendix}
\usepackage[caption=false,justification=justified]{subfig}

\newcommand{\beq}{\begin{equation}}
\newcommand{\eeq}{\end{equation}}
\newcommand{\bea}{\begin{eqnarray}}
\newcommand{\eea}{\end{eqnarray}}
\newcommand{\be}{\begin{equation}}      
\newcommand{\ee}{\end{equation}}

\def\nn{\nonumber}

\def\rightcontract{\mathop{\hbox{\vrule width0.5pt height6pt%
  \vrule height0.5pt width6pt}}}

\begin{document}

\title{Characterizing geodesic deviations in a Topological Star spacetime: massive, charged, spinning and stringy-like objects}

\author{Donato Bini$^{1}$, Giorgio Di Russo$^{2}$}
  \affiliation{
$^1$Istituto per le Applicazioni del Calcolo ``M. Picone,'' CNR, I-00185 Rome, Italy\\
$^2$School of Fundamental Physics and Mathematical Sciences, Hangzhou Institute for Advanced Study, UCAS, Hangzhou 310024, China\\
}

\date{\today}

\begin{abstract}
 We study deviations from geodesic motions in a Topological Star spacetime for either massive, charged  and spinning particles, elucidating  different behaviours with the Schwarzschild spacetime. We also consider the deviations for the motion of electrically charged stringy probes in $D=5$, framing all cases within a unified picture. 
\end{abstract}

\maketitle
\section{Introduction}

In the recent years the spacetime of a so-called Topological Star (TS), an exact solutions of the Einstein-Maxwell equations in five dimensions, has attracted the interest of many scientists, especially because of its remarkable difference with respect to the more familiar black hole (BH) case and, hence, has proven to be relevant to analyze the properties of a strong gravitational field from a different perspective with respect the BH one. For example,  the spacetime of a TS admits $1+4$ foliations with a quasi-Schwarzschild metric on the leaves of dimension $4$,  the existence of a spherical symmetry, the absence of horizons, the presence of an electromagnetic source, etc., see e.g. Refs.  \cite{Bah:2020ogh,Bah:2020pdz,Bah:2023ows}.

Most of the relevant kinematical properties of the TS spacetime have been already examined, like the spacetime geodesics \cite{Heidmann:2022ehn} (for which there exists a fully integrated form), waves propagation and background perturbations \cite{Bianchi:2023sfs,Heidmann:2023ojf,DiRusso:2024hmd,Cipriani:2024ygw,Bena:2024hoh}, and more recently also accelerated motions due to self-force effects \cite{Bianchi:2024vmi,Bianchi:2024rod,DiRusso:2025lip}. 

Moreover, TSs provide a fertile playground for studying the relevant physics of fuzzball-like objects \cite{Lunin:2001jy,Bena:2007kg,Skenderis:2008qn,Bianchi:2022qph} in a simplified scenario where the high degree of symmetry, combined with spherical symmetry, allows for the separation of radial and angular motion either at level of geodesics and for (massless) scalar field dynamics \cite{Chandrasekhar:1985kt}. A detailed study of certain symmetry properties of geodesic motion in D-brane and fuzzball geometries is available in \cite{Bianchi:2017sds,Bianchi:2021yqs,Bianchi:2022wku}, along with some machine learning applications to the integration of geodesic motion \cite{Cipriani:2025ini}.

Much is still to be examined. For example, very recently a rotating generalization of TSs was found \cite{Bianchi:2025uis}, stimulating comparisons between this new solution and the Kerr geometry.
Our aim here is to  study deviations from geodesic motion 1) of another geodesic which happens to be aligned with the first at a certain spacetime (initial) point; 2) of a family of particles endowed with a small mechanical spin; 3) of a magnetically charged particle; 4) of an electrically charged string. In all cases we aim at  characterizing (both from a geometrical and a physical point of view) the dynamics of    test bodies  orbiting around a TS  and to study the differences with respect to the Schwarzschild case. 

Furthermore, since the TS spacetime has an additional structure (i.e., it is characterized by two parameters $r_b$ and $r_s$, both  with dimensions of a length) with respect to the Schwarzschild spacetime (parametrized instead only by the Schwarzschild radius, $r_s$) it is natural to 
study the motion of particles endowed with an internal structure (the charge, the spin, etc.) in a TS background, in order to explore competing effects among different length scales. Indeed, deviations from geodesic motion will be 
expressed in terms of all such parameters and the spacetime parameters as well, and 
we will analyze here the corresponding interplay (in some smallness limit). 
Finally, we also include in the discussion the case of stringy charged probes in $D=5$: In this particular case we will contrast with classical Schwarzschild behaviors (now concerning the center-of-mass of the string itself).

Bunches (families) of particles can be chosen according to the dependence of their world lines on certain parameters (e.g., energy, angular momentum, spin, charge, etc.). Here, we will often consider (as an example) a LISA-inspired configuration of three particles, initially placed at the vertices of an ideal triangle (equilateral, for example), and then evolving with time, so that the triangle itself changes its shape during the evolution.  

Notation and conventions adopted here follow standard general relativity textbooks, e.g. \cite{Misner:1973prb}. We often work with units such that $G_N=1=c$ and choose the metric signature to be mostly positive.

\section{Topological Star metric}

The spacetime of a  TS is a  static and spherically symmetric manifold, and  an electro-vacuum solution in $D=5$ of the Einstein-Maxwell equations. It is  
described by the  metric
\bea\label{metric}
ds^2&=&-f_s(r)dt^2+\frac{dr^2}{f_s(r)f_b(r)}\nn\\
&+& r^2(d\theta^2+\sin^2\theta d\phi^2)+f_b(r)dy^2\,,
\eea
with 
\be
f_{s,b}(r)=1-\frac{r_{s,b}}{r},
\ee
in coordinates $(t,r,\theta,\phi,y)$ adapted to the staticity ($t$), the azimuthal symmetry ($\phi$) and to the extra dimension ($y$).
Moreover, the coordinate $y$ is periodic  $y\sim y+2\pi R_y$, with  $R_y=2r_b^{3/2}/(r_b-r_s)^{1/2}$. 
We will find it convenient to look at the metric \eqref{metric} as the superposition of a 2-sphere (with radius $r$)
\beq
ds^2_{(\theta,\phi)}=r^2(d\theta^2+\sin^2\theta d\phi^2)\,,
\eeq
and a 3-metric
\beq
\label{3-met}
ds_{(t,r,y)}^2=-f_s(r)dt^2+\frac{dr^2}{f_s(r)f_b(r)}+f_b(r)dy^2\,,
\eeq
such that $ds^2=ds_{(t,r,y)}^2+ds^2_{(\theta,\phi)}$.
Since the geometric properties of the 2-sphere are very well known, we summarize in Appendix \ref{dsquad_try} some geometrical information on $ds^2=ds_{(t,r,y)}^2$.
The metric source is an electromagnetic field, superposition of an electric 3-form field
\beq\label{elfieldstr}
F^{\rm (e)}=\frac{Q}{r^2} dr\wedge dt \wedge dy=d A^{\rm (e)}\,, 
\eeq
with
\beq\label{eqsource}
A^{\rm (e)}=-\frac{Q}{r}dt \wedge dy\,,
\eeq
and a magnetic 2-form field
\beq
F^{\rm (m)}=P \sin\theta  d\theta \wedge d\phi=d A^{\rm (m)}\,, 
\eeq
with
\beq
A^{\rm (m)}=-P \cos\theta d\phi\,,
\eeq
and where
\beq
P^2+Q^2=\frac{3r_b r_s}{2\kappa_5^2}\,.
\eeq
The corresponding electromagnetic energy-momentum tensor writes
\bea
T_{\mu\nu}=T_{\mu\nu}^{\rm (m)}+T_{\mu\nu}^{\rm (e)}\,,
\eea
where
\bea
T_{\mu\nu}^{\rm (m)}&=& F^{\rm (m)}_{\mu\alpha}F^{\rm (m)}_{\nu}{}^\beta -\frac14 g_{\mu\nu} F^{\rm (m)}_{\alpha\beta}F^{\rm (m)}{}^{\alpha\beta}\,, \nonumber\\
T_{\mu\nu}^{\rm (e)}&=& \frac12  \left[F^{\rm (e)}_{\mu\alpha\beta}F^{\rm (e)} _{\nu}{}^{\alpha\beta}-\frac16 g_{\mu\nu} F^{\rm (e)}_{\alpha\beta\gamma}F^{\rm (e)}{}^{\alpha\beta\gamma} \right]\,.\qquad
\eea
Finally, the Einstein-Maxwell's equations can be cast in the form
\bea
R_{\mu\nu}=\kappa_5^2 \left(T_{\mu\nu}-\frac13 g_{\mu\nu}T\right)\,,
\eea
with $T=T_{\mu}{}^\mu$ and  
with
\beq
8\pi G_4=\kappa_4^2=\frac{\kappa_5^2}{2\pi R_y}=\frac{8\pi G_5}{2\pi R_y}=\frac{4G_5}{R_y}\,.
\eeq

It is well known that  the metric \eqref{metric} is associated with two different regimes: 1) black string (BS): $r_b<r_s$, with an event horizon at $r=r_s$; 2) TS: $r_s<r_b$. Furthermore,   stability against (metric) linear perturbations requires $r_b<r_s<2r_b$
in the BS case and $r_s<r_b<2r_s$ in the TS one.
Here, unless differently indicated, we will mainly work with the case $Q=0$. In Section \ref{string} we will, however, restore the general case $Q\not=0$.

After dimensional reduction to $D=4$ (i.e., examined on the leaves $y=$constant of a natural $1+4$ foliation), the solution shows a naked singularity
and has a mass
\beq
G_4M_{\rm TS} = \frac{r_s}2  + \frac{r_b}4 \,. 
\eeq

Hereafter we will set $G_4\equiv G \equiv G_{\rm N}= 1 = c$. For $r_b = 0$, and
thus $r_s = 2GM_{\rm TS}$, the resulting singular solution is 
Schwarzschild BH$\times S_1$.
Eventually, we will use a Schwarzschild-inspired notation, i.e., denote
$r_s=2M$, $r_b=\alpha r_s=2\alpha M$,
with $M$ a common length scale, not to be confuse with $M_{\rm TS}$, the mass of the TS, unless for $r_b=0$. In fact,
\beq
M_{\rm TS} =\frac{r_s}{4}(\alpha+2)=\frac{M}{2}(\alpha+2) \,. 
\eeq
As stated above, for the TS spacetime the geodesics are well known \cite{Bah:2020ogh,Heidmann:2022ehn,Bah:2023ows,Bena:2024hoh} and we briefly resume only the results that we need in the following. The mass shell condition in Hamiltonian form reads 
\begin{equation}\label{massshellH}
\mathcal{H}=g^{\mu\nu}P_\mu P_\nu=-m^2\,,
\end{equation}
where $P_\mu$ are the conjugate canonical momenta. Since the metric is independent on $t$ and $\phi$, the associated momenta $P_t$ and $P_\phi$ are conserved along geodesics
\begin{equation}
    P_t=-E\,,\quad P_\phi=L\,.
\end{equation}
The solution of the Hamilton's equations is obtained by separation between the radial and the angular dynamics with the introduction of a Carter-like constant $K$
\bea\label{eqeffpot}
P_r^2&=&Q_R(r)=\frac{r^2E^2- f_s(r)(K^2+m^2 r^2)}{r^2 f_b(r)f_s(r)^2}\,,\nonumber\\
P_\theta^2&=& 
K^2-\frac{L^2}{\sin^2(\theta)}\,.
\eea
Because of the  spherical symmetry, we can study the equatorial motion $\theta=\pi/2$ without loss of generality,  i.e., assuming  $K=L$. Furthermore, we can rescale the energy and the angular momentum in units of the mass of the probe, see below. 
The circular  equatorial geodesics are defined by the conditions 
\begin{equation}
    Q_R(r_0,E_0,L_0)=0=Q_R'(r_0,E_0,L_0)\,,
\end{equation}
which can be solved as follows
\bea\label{critcond}
    \hat{E}_0&=&\frac{E_0}{m}=\frac{r_0-r_s}{r_0\sqrt{1-\frac{3 r_s}{2 r_0}}}\,,\nonumber\\ 
\hat{L}_0&=&\frac{L_0}{m}=\frac{\sqrt{r_0}\sqrt{r_s}}{\sqrt{2}\sqrt{1-\frac{3r_s}{2r_0}}}\,.
\eea
Let us introduce the notation
\beq
f_n(r)=1-n\frac{r_s}{r}\,,\qquad f_1(r)\equiv f_s(r)\,,
\eeq
largely used below. Here, $f_n(r)$ has the advantage of being a dimensionless quantity, and, for example
\begin{equation}\label{critcond2}
    \hat{E}_0= \frac{f_s(r_0)}{\sqrt{f_{3/2}(r_0)}}\,,\qquad \frac{\hat{L}_0}{r_s} =\frac{\sqrt{r_0 }}{\sqrt{2r_s}\sqrt{f_{3/2}(r_0)}}\,.
\end{equation}
Very recently, the orbits of scalar charges accelerated by a \lq\lq self force" have been analyzed too \cite{Bianchi:2024vmi,DiRusso:2025lip}.
Differently, the motion of extended bodies, charged particles or deformation effects due to tidal forces, has never been examined yet, and this is one of the goals of the present article.

We will start by studying geodesic deviations~\footnote{Geodesic deviation is (in a sense) a mathematical  formulation of what more generally is referred to as \lq\lq tidal forces." We recall that in the literature there exist variations on the definition of geodesic deviations, like \lq\lq connecting vectors" or tensors, relativistic strains, etc. \cite{Bini:2006ime,Bini:2007gxn,Bini:2007zzf,Bini:2007hd,Bini:2008zzb}}, i.e., 
analyzing  how nearby geodesics evolve relative to each other in presence  of gravitational forces generated by the field of a TS.
As it is well known, in the absence of external forces (including gravity), the relative velocity of any two geodesics is constant, whereas oscillations around/deviations from a reference geodesic characterize the behavior in a Schwarzschild spacetime. 
The TS spacetime is peculiar, since a cap replaces the horizon. While particles can fall into the BH horizon (and never escape), in the TS case the cap represents the boundary of the spacetime itself and is typically reached with a smooth tangential-like approach, as we will discuss below.

Extended  bodies (spinning bodies or, more in general,  bodies with an n-polar structure) or electrically charged particles do not follow geodesics.
In the literature force-driven motions are poorly studied with some exception like accelerated equatorial orbits around BH (where the circular symmetry simplifies their description).  
We will proceed here by examining some \lq\lq key" situations in which the structure of the body (spin, charge) competes with the additional structures (for example,  the additional scale which characterizes a TS  in comparison with the Schwarzschild spacetime), comparing and contrasting among them.
Explicit analytic (wherever the description will allow it) and numerical examples will be constructed on purpose, in order to distinguish between TS and Schwarzschild BHs.

\section{Geodesic deviation}

Let $U^\alpha(\tau)=\frac{dx^\alpha}{d\tau}$ be a generic timelike geodesic of the TS spacetime, parametrized by the proper time $\tau$ and characterized by   the energy and angular momentum parameters, say $U^\alpha(\tau; E,L)$.
Actually, because of the dependence on these parameters, $U^\alpha(\tau; E,L)$ identifies a family of geodesics, and after having selected a \lq\lq reference geodesic," say $U_*$ (corresponding to chosen values for $E_*$ and $L_*$) in the family, it is interesting to study how the other orbits will deviate from this one, The latter deviations  can be finally interpreted as a measure of the strength of TS gravitational field itself. 

To this end, according to a standard procedure, one introduces a deviation vector, say $\eta(\tau)$, defined all along the reference (timelike) geodesic and satisfying the geodesic deviation equation
\beq
\label{geo_dev_eq}
\frac{D^2 \eta^\alpha}{d\tau^2}+R^\alpha{}_{\beta\gamma\delta} U_*^\beta U_*^\delta \eta^\gamma \bigg|_{x^\alpha=x_*^\alpha(\tau)}=0\,.
\eeq 
Here 
\beq
\frac{D^2 \eta^\alpha}{d\tau^2}=\nabla_{U_*} (\nabla_{U_*} \eta^\alpha)\bigg|_{x^\alpha=x_*^\alpha(\tau)}\,,
\eeq
and 
\beq
{\mathcal E}(U_*)^\alpha{}_\gamma =R^\alpha{}_{\beta\gamma\delta} U_*^\beta U_*^\delta
\eeq
denotes the electric part of the Riemann tensor as measured by the observers $U_*^\alpha$. ${\mathcal E}(U_*)^\alpha{}_\gamma$ exists only along the $U_*$ world line and 
$\eta^\alpha(\tau)$ connects the reference geodesic to another one within the same family. Furthermore, one assumes $\eta^\alpha$ orthogonal to $U_*^\alpha$, since any component parallel to $U_*$ would have a trivial evolution, i.e., 
\beq
\label{orth_cond}
\eta\cdot U_*=0\,.
\eeq
Consequently,  only the evolution of three components of $\eta$ needs to be studied. 
Let us denote by $Y(\tau)$ the tangent vector to the deviation curve, say
\beq
Y^\alpha(\tau)=\frac{D\eta^\alpha(\tau)}{d \tau}\,,
\eeq
so that
\be
x(\tau)^\mu=x_*^\mu(\tau)+\eta^\mu(\tau)\,,
\ee
where, as stated above,  
\be\label{geodX}
x_*^\mu(\tau)=\left(t_p(\tau), r_p(\tau), \frac{\pi}{2}, \phi_p(\tau)\right)
\ee
describes a reference geodesic. More in detail our study concerns the following cases: 1) deviations from a circular equatorial geodesic and 2)  deviations from an unbound equatorial geodesics.
In the latter case we find it convenient to replace the energy per unit mass  in terms of velocity, $\hat E=\gamma=(1-v^2)^{-1/2}$, and the angular momentum per unit mass in terms of the  impact parameter $b$, namely $\hat L=b\gamma v$. The interest for these orbits is motivated  by recent applications of self force \cite{Bianchi:2024vmi,DiRusso:2025lip}.

%%%%%%%%%%%%%%%
\subsection{Deviations from a circular equatorial geodesic}

Let's start by studying the deviation   $\eta^\mu(\tau)$ from the (massive) circular reference geodesic at radius $r=r_0$ whose four velocity is $U_*=U_{\rm circ}$
\bea
\label{Ucirc}
U_{\rm circ}&=& 
\Gamma (\partial_t +\Omega \partial_\phi)\,,
\eea
with
\beq
\label{Gammadef}
\Gamma=f_{3/2}^{-1/2}(r_0)\,,\qquad \Omega=\sqrt{\frac{r_s}{2 r_0^3}}\,.
\eeq
The corresponding parametric equations read
\beq
t_{\rm circ}(\tau)=\Gamma \tau\,,\quad r_{\rm circ}=r_0\,,\quad \theta_{\rm circ}=\frac{\pi}{2}\,,\quad \phi_{\rm circ}=\Gamma\Omega \tau\,.
\eeq
When writing the   deviations equations \eqref{geo_dev_eq} can be convenient to refer to the frame components of the $\eta^\alpha(\tau)$, with respect to
an orthonormal frame adapted to observers at rest with respect to the coordinates
\bea
e_{\hat t}&=&\frac{1}{\sqrt{-g_{tt}}}\partial_t\,,\quad e_{\hat r}=\frac{1}{\sqrt{ g_{rr}}}\partial_r\,,\quad e_{\hat \theta}=\frac{1}{\sqrt{ g_{\theta\theta}}}\partial_\theta\,,\nonumber\\
e_{\hat \phi}&=&\frac{1}{\sqrt{ g_{\phi\phi}}}\partial_\phi\,,\quad e_{\hat y}=\frac{1}{\sqrt{ g_{yy}}}\partial_y\,,
\eea
namely, introducing all along the orbit \eqref{Ucirc} the rescaled quantities
\bea
\eta^t(\tau) &=& \frac{\eta^{\hat t}(\tau)}{\sqrt{f_s(r_0)}},\quad
\eta^r(\tau) =  \eta^{\hat r}(\tau)\sqrt{f_s(r_0)f_b(r_0)},\nonumber\\
\eta^\theta(\tau) &=& \frac{1}{r_0}  \eta^{\hat \theta}(\tau),\quad 
\eta^\phi(\tau) = \frac{1}{r_0}\hat \eta^{\hat \phi}(\tau),\nonumber\\   
\eta^y(\tau) &=& \frac{\eta^{\hat y}(\tau)}{\sqrt{f_b(r_0)}}\,. 
 \eea 
We find that the $y$-equation is trivial
\beq
{\eta^{\hat y}}''(\tau)=0\,,
\eeq
and will be ignored hereafter, 
while the $\theta$-equation decouples 
\beq
{\eta^{\hat \theta}}''(\tau)+\Gamma^2\Omega^2 \eta^{\hat \theta}(\tau)=0\,,
\eeq
and it is immediately solved
\beq
\eta^{\hat \theta}(\tau)=A\sin\left(\Gamma \Omega \tau\right)\,,\qquad \Gamma=(f_{3/2}(r_0))^{-1/2}
\eeq
where $A$ is and integration constant and we have assumed $\eta^\theta(0)=0$. 
The remaining equations write
\bea
&&{\eta^{\hat t}}''(\tau)+2r_0 \Omega^2 \Gamma \sqrt{f_b(r_0)}{\eta^{\hat r}}'(\tau)=0\,,\nonumber\\
&&{\eta^{\hat r}}''(\tau)-2\Gamma \Omega \sqrt{f_b(r_0)f_s(r_0)} {\eta^{\hat \phi}}'(\tau)\nonumber\\
&&\qquad +2r\Gamma \Omega^2 \sqrt{f_b(r_0)}{\eta^{\hat t}}'(\tau)\nonumber\\
&&\qquad -3f_b(r_0)f_s(r_0)\Gamma^2\Omega^2 {\eta^{\hat r}}(\tau)=0\,,\nonumber\\
&&{\eta^{\hat \phi}}''(\tau)+2\sqrt{f_b(r_0)f_s(r_0)}\Gamma\Omega {\eta^{\hat r}}'(\tau)=0\,.
\eea
The $t$ and $\phi$ equations can be integrated one times immediately as
\bea
&&{\eta^{\hat t}}'(\tau)+2r_0 \Omega^2 \Gamma \sqrt{f_b(r_0)}{\eta^{\hat r}}(\tau)=C_1\,,\nonumber\\
&&{\eta^{\hat \phi}}'(\tau)+2\sqrt{f_b(r_0)f_s(r_0)}\Gamma\Omega {\eta^{\hat r}}(\tau)=C_2\,,\qquad
\eea
and the integration constants can be set to zero, $C_1=0=C_2$ if one assumes ${\eta^{\hat r}}(0)=0$ and ${\eta^{\hat t}}'(0)={\eta^{\hat \phi}}'(0)$.
With this choice ($C_1=0=C_2$)
\bea
{\eta^{\hat t}} (\tau)&=& -2r_0 \Gamma \Omega^2 \sqrt{f_b(r_0)}\int_0^\tau d\tau' {\eta^{\hat r}}(\tau')\,,\nonumber\\
{\eta^{\hat \phi}}(\tau)&=&- 2\sqrt{f_b(r_0)f_s(r_0)}\Gamma\Omega \int_0^\tau d\tau' {\eta^{\hat r}}(\tau')\,.
\eea
If one does not choose the simplifying condition $C_1=0=C_2$ but keeps $C_1$ and $C_2$ generic then 
\bea
{\eta^{\hat t}} (\tau)&=& -2r_0 \Gamma \Omega^2 \sqrt{f_b(r_0)}\int_0^\tau d\tau' {\eta^{\hat r}}(\tau')\nonumber\\
&+&C_1\tau+{\eta^{\hat t}} (0) \,,\nonumber\\
{\eta^{\hat \phi}}(\tau)&=&- 2\sqrt{f_b(r_0)f_s(r_0)}\Gamma\Omega \int_0^\tau d\tau' {\eta^{\hat r}}(\tau')\nonumber\\
&+&C_2\tau+{\eta^{\hat \phi}}(0)\,,
\eea
and the  equation for ${\eta^{\hat r}}$ becomes
\bea
{\eta^{\hat r}}''(\tau)+\Omega_c^2 {\eta^{\hat r}} (\tau)&=& {\mathcal S}  \,, 
\eea
where
\beq
{\mathcal S}=-2\sqrt{f_b(r_0)}\Gamma \Omega (C_1 r\Omega-C_2 \sqrt{f_s(r_0)})\,,
\eeq
and
\bea
\Omega_c
&=& \Gamma\Omega \sqrt{ f_b(r_0)f_3(r_0)}\,,
\eea
is the TS-modification to the standard Schwarzschild epicyclic frequency,
\beq
\Omega_c=\Omega_c^{\rm Schw}\sqrt{f_b(r_0)}\,.
\eeq
The latter expression, defining a measurable quantity and characterizing an important difference with the Schwarzschild spacetime, can have observational consequences, in the sense that
\bea
\frac{\Delta \Omega_c}{\Omega_c^{\rm Schw}}&=&\frac{\Omega_c-\Omega_c^{\rm Schw}}{\Omega_c^{\rm Schw}}\nonumber\\
&=& \sqrt{f_b(r_0)}-1\nonumber\\
&=& \sqrt{1-\frac{\alpha r_s}{r_0}}-1\,,
\eea
and for large values of $r_0\gg r_s$ 
\bea
\frac{|\Delta \Omega_c|}{\Omega_c^{\rm Schw}}
&=& \frac12 \frac{\alpha r_s}{r_0}\,.
\eea
Therefore, an uncertainty in $|\Delta \Omega_c|/\Omega_c^{\rm Schw}$ can be read as a deviation of BH behavior in favour of the presence of a TS.

\begin{figure*}
    \centering
    \includegraphics[width=0.35\linewidth]{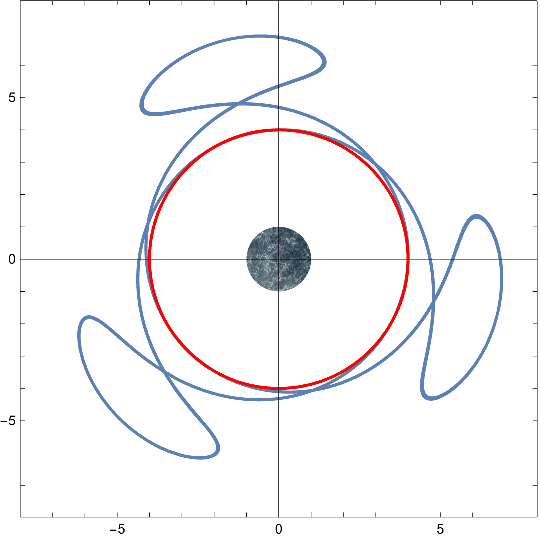}\qquad  
    \includegraphics[width=0.5\linewidth]{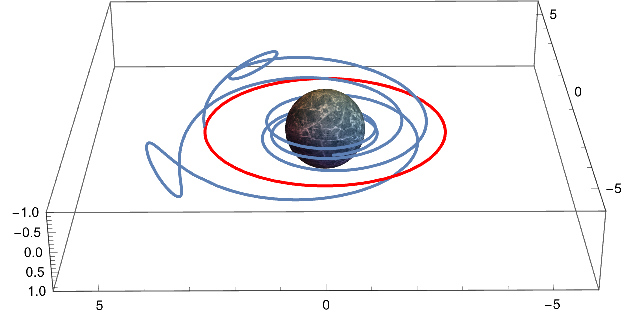}  
    \caption{Deviations from the circular geodesic motion for a TS. In the left panel the parameters are: $r_b=1$, $r_s=0.8$, $r_0=4$, $c_1=c_2=0.01$, $A=0$. For the right panel we set: $r_b=1$, $r_s=0.8$, $r_0=3$, $c_1=0.1$, $c_2=0$, $A=-0.1i$.}
    \label{circgeodeveq}
\end{figure*}

The solution for ${\eta^{\hat r}}(\tau)$ finally reads
\bea
{\eta^{\hat r}} (\tau)=\frac{{\mathcal S}}{\Omega_c^2}+A_1 \cos(\Omega_c\tau)+A_2 \sin(\Omega_c\tau)\,.
\eea
We can however restrict ourselves to the simple  case    ${\eta^{\hat r}}(0)={\eta^{\hat r}}'(0)=0$ and ${\eta^{\hat t}}(0)={\eta^{\hat \phi}}(0)=0$ (only these functions vanishing at $\tau=0$, and not their first derivatives).
This leads to the final solution
\bea
{\eta^{\hat r}} (\tau)=-\frac{{\mathcal S}}{\Omega_c^2} (1-\cos(\Omega_c\tau))\,,
\eea
with
\beq
\int_0^\tau d\tau' {\eta^{\hat r}} (\tau')=-\frac{{\mathcal S}}{\Omega_c^3}[\Omega_c \tau -\sin(\Omega_c\tau) ]
\eeq
which then give
\bea
{\eta^{\hat t}} (\tau)&=& 2r_0  \sqrt{f_b(r_0)}\Gamma \Omega^2\frac{{\mathcal S}}{\Omega_c^3}[\Omega_c \tau -\sin(\Omega_c\tau) ]\nonumber\\
&+& C_1\tau\,,\nonumber\\
{\eta^{\hat \phi}}(\tau)&=& 2\sqrt{f_b(r_0)f_s(r_0)}\Gamma\Omega  \frac{{\mathcal S}}{\Omega_c^3}[\Omega_c \tau -\sin(\Omega_c\tau) ]\nonumber\\
&+& C_2\tau\,.
\eea
A final remark concerns the orthogonality condition \eqref{orth_cond}, which translates as
\begin{equation}
\label{ortog_prop}
 \eta^{\hat t}(\tau)  =\frac{\Omega r_0}{\sqrt{f_s(r_0)}}   \eta^{\hat \phi}(\tau)\,,
\end{equation}
and can be used to relate $C_1$ and $C_2$, $C_1=\frac{\Omega r_0}{\sqrt{f_s(r_0)}} C_2$.
Plots of  equatorial as well as non equatorial geodesic deviations are shown in Fig. \ref{circgeodeveq}.

%%%%%%%%%%%%%%%%%%%%%%%%%
\subsection{Deviations from an unbound equatorial geodesics}

In equatorial case, unbound orbits  of the form   \eqref{geodX} are described by 
\bea
\frac{dt}{d\tau}&=&\frac{\hat{E}}{f_s(r)}\,,\nonumber\\
\frac{dr}{dt}&=&\frac{\sqrt{r-r_b}}{\hat{E}r^2}f_s(r)\sqrt{r^3(\hat{E}^2-1)+r_sr^2-\hat{L}^2(r-r_s)}\,,\nonumber\\
\frac{d\phi}{dt}&=&\frac{\hat{L}}{\hat{E}r^2}f_s(r)\,.
\eea
Let us use the following notation
\begin{equation}
    \hat{E}=\gamma=(1-v^2)^{-1/2},\quad \hat{L}=b\sqrt{\hat{E}^2-1}=b \gamma v\,,
\end{equation}
replacing energy in terms of velocity ($v$) and angular momentum in terms of impact parameter ($b$).
In the flat case, $r_s=r_b=0$, one finds  
\begin{equation}
    r|_{\rm flat}=\sqrt{b^2+v^2 t^2}\,.
\end{equation}
We will often use the dimensionless coordinate $T=\frac{v t}{b}$ in place of $\tau$, so that the covariant first and second derivatives of $\eta^\mu$ becomes
\bea
\frac{D\eta^\mu(T)}{d\tau}&=&u^t(T)\frac{v}{b}\Big[\frac{d\eta^\mu(T)}{dT}+\Gamma^{\mu}_{\nu\sigma}(T)\eta^\nu(T)\frac{d x_*^\sigma(T)}{dT}\Big]\,,\nonumber\\
\frac{D^2\eta^\mu(T)}{d\tau^2}&=&u^t(T)\frac{v}{b}\Big[\frac{d}{dT}\left(\frac{D\eta^\mu(T)}{d\tau}\right)\nonumber\\
&+&\Gamma^\mu_{\nu\sigma}(T)\frac{D\eta^\nu(T)}{d\tau}\frac{dx_*^\sigma(T)}{dT}\Big]\,,
\eea
with $u^t(T)=\frac{dt(T)}{d\tau}$. 
Note that both the Christoffel symbols and Riemann tensor need to be evaluated along the reference geodesic  orbit, and hence have been denoted as functions of $T$ too.

Passing to the dimensionless time parameter $T=vt/b$ and introducing the dimensionless Post-Minkowskian (PM) parameter 
\be
\epsilon=\frac{GM}{b v^2}=\frac{Gr_s}{2 b v^2}\,,
\ee
i.e., $\epsilon=\frac{r_s}{2 b v^2}$ when using $G=1$, we can write the solution of the geodesic equations in the following PM-expanded form
\bea
r_p(T)&=& b\sqrt{1+T^2}+r_1(T)\epsilon+ r_2(T)\epsilon^2+O(\epsilon^3)\,,\nonumber\\
\phi_p(T)&=& {\rm arctan}(T) +\epsilon \phi_1(T)+\epsilon^2 \phi_2(T)+O(\epsilon^3) \,,\nonumber\\
u^t(T)&=& \gamma+u_1^{t}(T)\epsilon+u_2^{t}(T)\epsilon^2+O(\epsilon^3)\,,
\eea
where the various coefficients are conveniently collected in Table \ref{tab:ri-phii-uti} using a compact notation summarized in Table \ref{tab:notation}.
\begin{table*}  
\caption{\label{tab:ri-phii-uti} List of the various coefficients entering the PM expansion of the unbound geodesics in the TS spacetime, in powers of $\epsilon=\frac{r_s}{2bv^2}$.
Note that $r(0)=r_{\rm min}$, i.e., $T=0$ corresponds to the turning point.
}
\begin{ruledtabular}
\begin{tabular}{ll}
$\frac{r_1}{b}$ & $ -1+\tilde f^{\rm as1}_{1/2}-v^2(\alpha +3) \tilde f^{\rm as1}_{1/2}$\\
$\frac{r_2}{b}$ & $\frac{1}{2} f_{1/2}  +\tilde f^{\rm as1}_{1} 
+\frac12 f^{\rm as2}_{3/2}+v^2\Big[-2 f_{1/2}
-2(\alpha +3)\tilde f^{\rm as1}_{1} -(\alpha +3) f^{\rm as2}_{3/2} \Big]$\\
   $$ & $+v^4\Big[ -\frac{3}{2} \left(\alpha ^2+2 \alpha +5\right) \tilde f^{\rm at1}_{1/2}  +(\alpha +3)^2 \tilde f^{\rm as1}_1
+\frac12 (\alpha +3)^2 f^{\rm as2}_{3/2}\Big]$\\
 \hline
$\phi_1$ & $\tilde f_{1/2}+f^{\rm as1}_1 +v^2\Big[(\alpha +1)\tilde f_{1/2}  -(\alpha +3)f^{\rm as1}_1  \Big]$\\
$\phi_2$ & $2 f^{\rm as1}_{3/2}-\tilde f^{\rm as2}_2+v^2\Big[(\alpha +3) \tilde f_1 +2 (\alpha +3)\tilde f^{\rm as2}_2 
-2 (\alpha +4)f^{\rm as1}_{3/2}+(\alpha +3)\left(f^{\rm at0}-\frac{\pi}{2}\right)  \Big] $\\
&$v^4 \Big[\frac14 \left(3 \alpha ^2+2 \alpha +3\right) \tilde f_1 
-\frac{3}{2}\left(\alpha ^2+2 \alpha +5\right)f^{\rm at1}_1 
-(\alpha +3)^2 \tilde f^{\rm as2}_2  
+2 (\alpha +3) f^{\rm as1}_{3/2}+
\frac{1}{4} \left(3 \alpha ^2+2\alpha +3\right) \left(f^{\rm at0}-\frac{\pi}{2}\right)\Big]$\\
\hline
$u^t_1$ &$2 v^2f_{1/2}+v^4f_{1/2}$\\
$u^t_2$ & $ 2v^2(f_1-\tilde f^{\rm as1}_{3/2})+v^4(5f_1+(5+2\alpha))\tilde f^{\rm as1}_{3/2}$\\
\end{tabular}
\end{ruledtabular}
\end{table*}
The equations for deviations can also be solved in PM sense. For example, the first-order values of the displacement  are of the type
\bea
\eta_0^t(T)&=&A_{0,0}bT\,,\nonumber\\
\eta_0^r(T)&=&b\left(A_{3,0}f_{-\frac12}-A_{3,0} f_{\frac{1}{2}}+A_{1,0} \tilde{f}_{\frac{1}{2}}\right)\,,\nonumber\\
\eta_0^\theta(T)&=&A_{2,0}\tilde{f}_{\frac{1}{2}}\,,\nonumber\\
\eta_0^\phi(T)&=&A_{1,0}f_1+A_{3,0}\tilde{f}_1-A_{1,0}\,, 
\eea
and at the first order we can distinguish the TS $\alpha$ corrections from the Schwarzschild results
\begin{equation}
    \eta^\mu_{1}(T)=\eta^\mu_{1,{\rm Schw}}(T)+\alpha \Delta \eta^\mu_{1}(T)
\end{equation}
\begin{widetext}
\bea
\eta_{1,{\rm Schw}}^t(T)&=&A_{0,1}T+2  v A_{1,0}f_{\frac{1}{2}}+2 v A_{3,0}\tilde{f}_{\frac{1}{2}}-2v  (A_{1,0}+ A_{3,0}{\rm arcsinh}(T))\,,\nonumber\\
\eta_{1,{\rm Schw}}^r(T)&=&b\Big[3 A_{3,0}+A_{3,1}f_{-\frac12}-2A_{0,0}v-A_{3,0}v^2+(2A_{0,0}v-2A_{3,0}-A_{3,1})f_{\frac{1}{2}}+A_{1,1}\tilde{f}_{\frac{1}{2}} \nonumber\\
&+&(A_{3,0}v^2-A_{3,0})f_1+(A_{1,0}-A_{1,0}v^2)\tilde{f}_1+2(A_{0,0}v-A_{3,0}v^2-A_{3,0})\tilde{f}^{\rm as1}_{\frac{1}{2}}\nonumber\\
&+&(A_{1,0}-3A_{1,0}v^2)f^{\rm as1}_{\frac{3}{2}}+(A_{3,0}-3A_{3,0}v^2)\tilde{f}^{\rm as1}_{\frac{3}{2}}\Big]\,,\nonumber\\
\eta_{1,{\rm Schw}}^\theta(T)&=&A_{2,1}\tilde{f}_{\frac{1}{2}}+A_{2,0}(1+v^2)\tilde{f}_1+A_{2,0}(1-3v^2)f^{\rm as1}_{\frac{3}{2}}\,,\nonumber\\
\eta_{1,{\rm Schw}}^\phi(T)&=&-2A_{1,0}-A_{1,1}+2A_{1,0}f_{\frac{3}{2}}+2A_{3,0}\tilde{f}_{\frac{3}{2}}+A_{1,1}f_1+(2A_{3,0}+A_{3,1}-2A_{0,0}v)\tilde{f}_{1}+2(A_{0,0}v-A_{3,0})\tilde{f}_{\frac{1}{2}}\nonumber\\
&+&2(A_{0,0}v-2A_{3,0}+2A_{3,0}v^2)f^{\rm as1}_1+2(A_{3,0}-3A_{3,0}v^2)f^{\rm as1}_{2}\nonumber\\
&+&2(3A_{1,0}v^2-A_{1,0})\tilde{f}^{\rm as1}_2\,.
\eea
\bea
\Delta\eta_1^t(T)&=&0\nonumber\\
\Delta\eta_1^r(T)&=& v^2\Big[A_{3,0}(-1+ f_1-\tilde{f}^{\rm as1}_{\frac{3}{2}})-A_{1,0}(\tilde{f}_1+ f^{\rm as1}_{\frac{3}{2}})\Big]\,,\nonumber\\
\Delta\eta_1^\theta(T)&=& v^2 A_{2,0} \left(\tilde{f}_1- f^{\rm as1}_{\frac{3}{2}}\right)\,,\nonumber\\
\Delta\eta_1^\phi(T)&=&2v^2 \Big[A_{3,0} (f^{\rm as1}_1- f^{\rm as1}_{2})+A_{1,0}\tilde{f}^{\rm as1}_2\Big]\,.
\eea
\end{widetext}
Concerning the notation introduced for the constants of integration $A_{i,j}$, the first index refers to the component of the displacement vector $\eta^\mu(T)$ while the second recalls the PM order where it appears; the various functions, instead, are defined according Table \ref{tab:notation}.

\begin{table}  
\caption{\label{tab:notation}  Useful compact notation.
}
\begin{ruledtabular}
\begin{tabular}{ll}
${{f}}_n$ & $\frac{1}{(T^2+1)^n}$\\
$\tilde{{f}}_n$ & $\frac{T}{(T^2+1)^n}$\\
${{f}}_n^{\rm ask}$ & $\frac{{\rm arcsinh}^k(T)}{(T^2+1)^n}$\\
$\tilde f_n^{\rm ask}$ & $\frac{T{\rm arcsinh}^k(T)}{(T^2+1)^n}$\\
%%%%
${{f}}^{\rm at0}$ & ${\rm arctan}(T)+\frac{\pi}{2}$\\
${{f}}_n^{\rm at1}$ & $\frac{{\rm arctan}(T)}{(T^2+1)^n}$\\
$\tilde{{f}}_n^{\rm at1}$ & $\frac{T{\rm arctan}(T)}{(T^2+1)^n}$\\
${{f}}_n^{\rm log}$ & $\frac{{\rm Log}(1+T^2)}{2(1+T^2)^n}$\\
 \hline
\end{tabular}
\end{ruledtabular}
\end{table}
The orthogonality condition \eqref{orth_cond} implies the following relations between the integration constants
\be
A_{0,0}=A_{3,0}v,\quad A_{0,1}=A_{3,1} v, \quad A_{0,2}=A_{3,2} v\,.
\ee
 In principle the integration constants $A_{i,j}$ are 24, but the orthogonality condition fixes 3 of these constants and the requirement that the displacement components vanish in $T=0$ provide other 12 constraints (four constraints for each $\epsilon$ order considered, three in our case), so that we are left with only 9 independent constants.
This means that the only unconstrained integration constants are $A_{i,j}$ for $i=1,2,3$ and $j=0,1,2$. The equatorial case corresponds to $A_{2,j}=0$ for $j=0,1,2$. 
A simple, equatorial, solution can be given, for example, by setting  $A_{2,0}=A_{2,1}=0$ as well $A_{0,1}=0$, $A_{1,0}=0$, $A_{1,1}=0$, $A_{3,1}=0$,  $A_{0,0}=A_{3,0}v$ (from the orthogonality condition) and finally $A_{3,0}=1$
\bea
\eta_0^t(T)&=& vbT\,,\nonumber\\
\eta_0^r(T)&=& b\left(f_{-\frac{1}{2}}- f_{\frac{1}{2}}\right)\,,\nonumber\\
\eta_0^\theta(T)&=&0\,,\nonumber\\
\eta_0^\phi(T)&=& \tilde{f}_1\,,\nonumber\\ 
\eta_1^t(T)&=&b v\left(2   \tilde{f}_{\frac{1}{2}}-2\,f^{\rm as1}_0 \right)\,,\nonumber\\
\eta_1^r(T)&=&b\Big[3 (1 -v^2)-v^2\alpha-2(1-v^2)f_{\frac{1}{2}}\nonumber\\
&-&(1-v^2(1+\alpha))f_1-2\tilde{f}^{\rm as1}_{\frac{1}{2}}\nonumber\\
&+& (1-v^2(3+\alpha))\tilde{f}^{\rm as1}_{\frac{3}{2}}\Big]\,,\nonumber\\
\eta_1^\theta(T)&=&0\,,\nonumber\\
\eta_1^\phi(T)&=&  2\tilde{f}_{\frac{3}{2}}+2( 1-v^2)\tilde{f}_{1}-2(1-v^2)\tilde{f}_{\frac{1}{2}}\nonumber\\
&+&2((3+\alpha)v^2-2)f^{\rm as1}_1+2(1-v^2(3-\alpha))f^{\rm as1}_{2}\,.\nonumber\\
\eea
The presence of all such constants in the general solution, however, complicates any explicit comparison with the Schwarzschild spacetime. For example, the radial component of the deviation differ from the corresponding quantity in  the Schwarzschild spacetime by first-order in $\epsilon$ modifications
\bea
\delta \eta^r(T)&=& \eta^r(T)-\eta_{\rm Schw}^r(T)\nonumber\\
&=& -bv^2\alpha\epsilon\Big[A_{3,0}(1-f_1+\tilde{f}^{\rm as1}_\frac{3}{2})\nonumber\\
&+&A_{1,0}(\tilde{f}_1+f^{\rm as1}_{\frac{3}{2}})\Big]+\mathcal{O}(\epsilon^2)\,,
\eea
and then (restoring the factor $\epsilon$ of the PM expansion)
\bea
\frac{\Delta \eta_1^r(T)}{\eta_{1\rm Schw}^r(T)}&=& - v^2\alpha \epsilon\, {\mathcal K}(T) \,,
\eea
where
\bea
{\mathcal K}(T)&=& f_{\frac{1}{2}}+\frac{{\rm arcsinh}(T)}{T}-\tilde{f}^{\rm as1}_1\,,\qquad
\eea
which varies from $0$ to $2$ for all $T\in {\mathbb R}$ (with maximum value ${\mathcal K}(0)=2$ and showing a bell-shaped trend).
Therefore, the maximum of the radial deviation is given by
\beq
\frac{\Delta \eta_1^r(T)}{\eta_{1\rm Schw}^r(T)}\sim  2 v^2 \alpha \epsilon\,,
\eeq
which again can be possibly compared with observations, in the sense that  \lq\lq error bars" associated with the determination of such a quantity might be compatible with  the presence of certain, nonzero value $\alpha$, i.e., might imply that the considered object is more likely  a TS rather than a Schwarzschild BH.

A different approach, also valid in the non-geodesic case, i.e., for orbits accelerated with acceleration $a(U)=\nabla_{U} U$ (hereafter  \lq\lq strain approach" \cite{Bini:2006ime, Bini:2007gxn,Bini:2007zzf,Bini:2007hd}) consists in introducing along the 
world lines  of the congruence $U$ a connecting vector, $Y$ defined so that
\beq
\label{eq:1}
\pounds_{U} Y=0\  \qquad \rightarrow \qquad \nabla_{U}Y=\nabla_YU\,,
\eeq
for all the world lines in the congruence, 
which becomes 
\beq
\label{eq_Y_1stord}
\frac{D Y}{d \tau}=\nabla_Y U= -(Y\cdot U)a(U)-K(U)\rightcontract Y\,,
\eeq
after introducing the kinematical tensor $K(U)$, defined  from the  covariant derivative of $U$, i.e.
\beq
\nabla_\alpha U^\beta =-U_\alpha a(U)^\beta -K(U)^\beta{}_\alpha\,.
\eeq
The symbol $\rightcontract $ stands for the right-contraction operation among tensors (the rightmost covariant index of the first tensor in the product is contracted with the uppermost contravariant index of the second tensor).

\section{Spinning particles}

Consider an extended body endowed with a spin structure. 
The Mathisson-Papapetrou-Dixon (MPD) equations \cite{math37,papa51,tulc59,dixon64,dixon69,dixon70,dixon73,dixon74,ehlers77}
 regulate the orbit and the transport law along it of the dipolar (spin) structure
\begin{eqnarray}
\label{papcoreqs1}
\frac{{\rm D}P^{\mu}}{d \tau} & = &
- \frac12 \, R^\mu{}_{\nu \alpha \beta} \, U^\nu \, S^{\alpha \beta} \,,
\\
\label{papcoreqs2}
\frac{{\rm D}S^{\mu\nu}}{d \tau} & = & 
2 \, P^{[\mu}U^{\nu]} \,,
\end{eqnarray}
where $P^{\mu}=m u^\mu$ (with $u \cdot u = -1$) is the total 4-momentum of the body with mass $m$, $S^{\mu \nu}$ is a (antisymmetric) spin tensor, and $U^\mu=d z^\mu/d\tau$ is the timelike unit tangent vector of the \lq\lq center-of-mass-line'' (with parametric equations $x^\mu=z^\mu(\tau)$) used to make the multipole reduction, parametrized by the proper time $\tau$, see also Refs. \cite{quadrup_schw,quadrup_kerr1,quadrup_kerr_num,spin_dev_schw}. 

In order the model to be mathematically consistent the following additional conditions\footnote{It is  customary to call these conditions as Covariant Spin Conditions (CSC). However, there exist different choices (also covariant) used in the literature \cite{Bini:2000vv,Bini:2011nhv,Bini:2011tvf}.} should be imposed \cite{tulc59,dixon64}
\beq
\label{tulczconds}
S^{\mu\nu}u{}_\nu=0\,.
\eeq
Consequently, the spin tensor can be fully represented by a spatial (with respect to $u$) vector,
\beq
S(u)^\alpha=\frac12 \eta(u)^\alpha{}_{\beta\gamma}S^{\beta\gamma}
\,,
\eeq
where $\eta(u)_{\alpha\beta\gamma}=\eta_{\mu\alpha\beta\gamma}u^\mu$ is the spatial (with respect to $u$) unit volume 3-form with $\eta_{\alpha\beta\gamma\delta}=\sqrt{-g} \epsilon_{\alpha\beta\gamma\delta}$ the unit volume 4-form and $\epsilon_{\alpha\beta\gamma\delta}$ ($\epsilon_{0123}=1$) the Levi-Civita alternating symbol. 
It is also useful to introduce the signed magnitude $s$ of the spin vector
\beq
\label{sinv}
s^2=S(u)^\beta S(u)_\beta = \frac12 S_{\mu\nu}S^{\mu\nu}\,, 
\eeq
which in general  is  not constant along the trajectory of the extended body. 

Here, we consider the center line of the type
\beq
U^\mu(\tau) =U^\mu_{\rm circ}(\tau)+\hat s  Y^\mu(\tau)\,,
\eeq
where 
\beq
\hat s= \frac{s}{m r_s}\,,
\eeq
is a natural dimensionless spin variable, 
$U^\mu_{\rm circ}(\tau)$ is an equatorial, timelike, circular geodesic at the coordinate radius $r=r_0$, already used in Eq. \eqref{Ucirc}
\beq
U_{\rm circ}(\tau)=\Gamma (\partial_t +\Omega \partial_\phi)\,,
\eeq
with $\Gamma=f_{3/2}^{-1/2}(r_0)$ and $\Omega=\sqrt{\frac{r_s}{2 r_0^3}}$
which always exist for $r_0>\frac{3}{2}r_s$ (at $r_0=\frac{3}{2}r_s$ the geodesic is no more timelike but becomes null).
Note that 1) $\Gamma$ does not depend on $r_b$, and hence the four velocity of the circular geodesic  writes exactly as in the Schwarzschild case; 2) because of the MPD construction as an expansion in multipoles around a reference world line, the model implicitly requires to be linear in spin. Indeed, spin-squared terms are of quadrupolar origin and should be inclued in the MPD model according to a well established procedure. One can anyway consider with certain caution quadratic terms in spin when, for example, an exact solution of the equations of motion can be obtained. This, however, can only bear an incomplete information. We therefore limit our consideartions below to the linear-in-spin case.
 
The world line of $U$ is then described by the following parametric equations
\beq
x^\mu(\tau)=x^\mu_{\rm circ}(\tau)+\hat{s}\eta^\mu(\tau)\,,
\eeq
with $t_{\rm circ}(\tau)=\Gamma \tau$, $r_{\rm circ}(\tau)=r_0$, $\theta_{\rm circ}(\tau)=\frac{\pi}{2}$, $\phi_{\rm circ}(\tau)=\Gamma\Omega \tau$ (the condition $y_{\rm circ}(\tau)=0$ makes trivial the discussion of any $y$ component in the dynamics, and hence we will ignore it) and
\beq
Y(\tau)=\frac{D\eta}{d\tau}\,.
\eeq
We choose 
the spin vector orthogonal to the equatorial (orbital) plane,
\beq
S(u)^\mu=-s e_{\hat \theta}=-\frac{s}{r_0} \partial_\theta \,,
\eeq
implying that the spin evolution equations are identically satisfied assuming $s$=constant and $P=m U$, i.e., $u=U+O(s^2)$.
We find
\be
S^{01}=-r_0   \sqrt{f_b(r_0)}\Gamma \Omega\, s\,,\qquad 
S^{13}=\frac{\sqrt{f_b(r_0)} f_s(r_0)}{r_0}\Gamma s\,,
\ee
Therefore,
\bea
\frac{d (m U^{\mu})}{d \tau} +\Gamma^\mu{}_{\alpha\beta}\big|_{x^\mu=x^\mu_{\rm circ}+\hat s \eta^\mu}m U^\alpha U^\beta  &=&  
F_{\rm spin}^\mu\,,\qquad
\eea
where one can take
\bea
F_{\rm spin}^\mu 
&=& - \frac12 \, R^\mu{}_{\nu \alpha \beta} \, U_{\rm circ}^\nu \, \, S^{\alpha\beta}\big|_{x^\mu=x^\mu_{\rm circ}}\,.
\eea
Consequently, because of the orthogonality with $U$ of both $F_{\rm spin}^\mu$ and $a(U)^\mu=\nabla_UU^\mu$
\beq
\frac{dm}{d\tau}=0\,,\quad \to \quad m={\rm const.}\,,
\eeq
and hence the motion equations reduce to
\bea\label{systspin}
 \frac{d  U^{\mu} }{d \tau} +\Gamma^\mu{}_{\alpha\beta}\big|_{x^\mu=x^\mu_{\rm circ}+\hat s  \eta^\mu} U^\alpha U^\beta  &=&  
\frac{1}{m}F_{\rm spin}^\mu\,.\qquad
\eea
From the constraint $U\cdot U=-1$, we obtain the relation
\bea
\label{orto_nonframe}
\eta^t(\tau)
&=& \frac{\Omega r_0^2 }{f_s(r_0)} \eta^\phi(\tau ) \,,
\eea
with $\Omega$ the geodesic angular velocity. In terms of frame components Eq. \eqref{orto_nonframe} reduces to Eq.  \eqref{ortog_prop} above.

Apart from the $\theta$ equation, which does not mix other components and writes
\beq
\frac{d^2\eta^\theta(\tau)}{d\tau^2}+\Gamma^2 \Omega^2 \eta^\theta(\tau)=0\,,
\eeq
and admits  after imposing $\eta^\theta(0)=0$, the solution 
\beq
\eta^\theta(\tau)=A \sin (\Gamma \Omega \tau)\,,
\eeq
as above, 
the radial and azimuthal components form a system of coupled ODE. Symbolically,
\bea
\label{eqspin2}
&&\frac{d^2\eta^{\phi}(\tau)}{d\tau^2}+B_1\frac{d\eta^r(\tau)}{d\tau}-\Sigma_\phi^2 \eta^{\phi}(\tau)=0\,,\nonumber\\
&& \frac{d^2\eta^r(\tau)}{d\tau^2}+B_2\frac{d\eta^\phi(\tau)}{d\tau}-\Sigma_r^2 \eta^r(\tau)+\bar{B}=0\,,
\eea
with
\bea
[B_1,B_2]
&=& 3\Omega \left[ \frac{1}{ r_0f_{3/2}^{1/2}(r_0)},-  r_0 f_b(r_0)f_{3/2}^{1/2}(r_0)  \right]\,,\nonumber\\
{}[\Sigma_r^2, \Sigma_\phi^2 ]
&=&\Omega^2f_b(r_0) \left[  \frac{5f_{6/5}(r_0) }{f_{3/2}(r_0)},  2 \right]\,, \nonumber\\
\bar{B}
&=&  -\Omega^3 \frac{r_0^2 \sqrt{f_b(r_0)}f_s(r_0)}{2f_{3/2}(r_0)}
\left(2   \frac{r_b}{r_0}+6\frac{r_s}{r_0} -9 \frac{r_b}{r_0} \frac{r_s}{r_0}\right)\,,\nonumber\\
\eea
so that
\begin{equation}
    B_1 B_2=-\frac{9 f_b(r_0) r_s}{2r_0^3}=- 9 f_b(r_0) \Omega^2=-\frac{9}{2}\Sigma_\phi^2\,.
\end{equation}
Note that if one uses frame components instead of coordinate components the only difference is a rescaling of the various $B_i$ coefficients into some $B_i^{\rm new}$ values, namely 
\bea
B_1^{\rm new}&=& r_0 \sqrt{f_s(r_0)f_b(r_0)} B_1\,,\nonumber\\
B_2^{\rm new}&=& \frac{B_2}{r_0 \sqrt{f_s(r_0)f_b(r_0)}}\,,\nonumber\\
\bar{B}^{\rm new}&=& \frac{\bar{B}}{\sqrt{f_s(r_0)f_b(r_0)}}\,,
\eea
and hence not offering any special advantage.

An equilibrium (constant) solution is located at
\beq
\eta^{\phi}_{\rm equil}(\tau)=0\,,\qquad
\eta^r_{\rm equil}=\frac{\bar{B}}{\Sigma_r^2}\,,\qquad \eta^{\theta}_{\rm equil}(\tau)=0\,.
\eeq
As a consequence,  writing $\eta^\alpha(\tau)=\eta^{\alpha}_{\rm equil}(\tau)+\bar \eta^{\alpha}(\tau)$ the new equations read
\bea
\label{sys_phi_r}
&&\frac{d^2\bar\eta^{\phi}(\tau)}{d\tau^2}+B_1\frac{d\bar\eta^r(\tau)}{d\tau}-\Sigma_\phi^2 \bar \eta^{\phi}(\tau)=0\,,\nonumber\\
&& \frac{d^2\bar\eta^r(\tau)}{d\tau^2}+B_2\frac{d\bar\eta^\phi(\tau)}{d\tau}-\Sigma_r^2 \bar\eta^r(\tau)=0\,,
\eea
and, for example, imply
\bea
&&B_2 \frac{d\bar \eta^\phi(\tau)}{d\tau} \left( \frac{d^2\bar \eta^{\phi}(\tau)}{d\tau^2}-\Sigma_\phi^2\bar \eta^{\phi}(\tau)\right)\nonumber\\
&-&B_1\frac{d\bar \eta^r(\tau)}{d\tau}\left(\frac{d^2\bar \eta^r(\tau)}{d\tau^2}-\Sigma_r^2\bar \eta^r(\tau)\right)=0\,,
\eea
namely the following conservation law
\beq
B_2{\mathcal E}_{\phi}(\tau)-B_1 {\mathcal E}_{r}(\tau)={\rm constant}\,,
\eeq
where the \lq\lq energy-like" variables are defined as
\bea
{\mathcal E}_{\phi}(\tau)&=&\frac12 \left( \frac{d\bar \eta^\phi(\tau)}{d\tau} \right)^2-\frac12 \Sigma_\phi^2(\bar \eta^\phi(\tau))^2\,,\nonumber\\
{\mathcal E}_{r}(\tau)&=& \frac12 \left( \frac{d\bar \eta^r(\tau)}{d\tau} \right)^2-\frac12 \Sigma_r^2(\bar \eta^r(\tau))^2\,. 
\eea
In spite of the apparent simplicity the system of coupled equations \eqref{sys_phi_r} requires to be studied with some care.

One possibility is the following.
Differentiating with respect to $\tau$ the first equation and injecting in it the second one gives
\bea
\label{sys_phi_r2}
&&\frac{d^3\bar\eta^{\phi}(\tau)}{d\tau^3}-(B_1B_2+\Sigma_\phi^2)\frac{d\bar\eta^r(\tau)}{d\tau}+B_1\Sigma_r^2 \bar \eta^{r}(\tau)=0\,,\nonumber\\
&& \frac{d^3\bar\eta^r(\tau)}{d\tau^2}-(B_1B_2+\Sigma_r^2)\frac{d\bar\eta^r(\tau)}{d\tau}+B_2\Sigma_\phi^2 \bar \eta^{\phi}(\tau)=0\,,\nonumber\\
\eea
where
\bea
(B_1B_2+\Sigma_\phi^2)&=& -\frac{7}{2}\Sigma_\phi^2\,,\nonumber\\
(B_1B_2+\Sigma_r^2)&=& -\frac{9}{2}\Sigma_\phi^2+\Sigma_r^2\,.
\eea
One can then isolate $\bar \eta^{r}(\tau)$ from the first equation of \eqref{sys_phi_r2} and substitute into the second one (and similarly for the other equation), obtaining the following (sixth-order) main equation 
\begin{widetext}
\bea
\label{eqsixthorder}
&&\frac{d^6X(\tau)}{d\tau^6} + \left(8\Sigma_\phi^2-\Sigma_r^2\right)\frac{d^4X(\tau)}{d\tau^4}+\frac74 \Sigma_\phi^2\left(9\Sigma_\phi^2-2\Sigma_r^2\right)\frac{d^2X(\tau)}{d\tau^2}+\frac92 \Sigma_\phi^4\Sigma_r^2 X(\tau)=0\,, 
\eea
\end{widetext}
valid (remarkably) for both $X=\bar\eta^{\phi}$ and $X=\bar\eta^{r}$.
Eq. \eqref{eqsixthorder} can be cast in the operatorial form
\beq
D_0(D_-(D_+ (X(\tau))))=0\,,
\eeq
where
\bea
D_0 &=&\frac{d^2}{d\tau^2}+\Omega_0^2\,, \nonumber\\
D_\pm &=&\frac{d^2}{d\tau^2}-\Omega_\pm^2\,, \nonumber\\
\eea
where~\footnote{One can formally perform further simplifications; for example,  in the region where ${\mathcal A}>0$
$$
\Omega_\pm = \frac{\sqrt{{\mathcal A}\pm\sqrt{{\mathcal D}}}}{2}=\frac12\left[ \sqrt{\frac{{\mathcal A}+4\Sigma_r \Sigma_\phi}{2}}\pm  \sqrt{\frac{{\mathcal A}-4\Sigma_r \Sigma_\phi}{2}} \right]\,.
$$
}
\bea
\Omega_0&=& \frac{3}{\sqrt{2}}\Sigma_\phi\,,\nonumber\\
\Omega_\pm^2&=& \frac{{\mathcal A}\pm\sqrt{{\mathcal D}}}{4}\,,\nonumber\\
{\mathcal A}&=& -7\Sigma_\phi^2+2\Sigma_r^2\,,\nonumber\\
{\mathcal D}&=&4\Sigma_r^4-44\Sigma_r^2\Sigma_\phi^2+49\Sigma_\phi^4\nonumber\\
&=& -9 (\Gamma \Omega)^4 f_b(r_0)^2\left(16 -40 \frac{r_s}{r_0} +23 \frac{r_s^2}{r_0^2}\right)\,,\nonumber\\
{\mathcal A}^2-{\mathcal D}&=& 16\Sigma_r^2\Sigma_\phi^2\,,
\eea

The solution of Eq. \eqref{eqsixthorder} involves then six constants and writes
\bea
\label{gensol}
X(\tau)&=& C_1 e^{-i\Omega_0\tau}+C_2 e^{i\Omega_0\tau}\nonumber\\
&+& C_3 e^{-\Omega_- \tau}+ C_4 e^{\Omega_- \tau}\nonumber\\
&+& C_5 e^{-\Omega_+ \tau}+ C_6 e^{\Omega_+ \tau}\,,
\eea 
where the constants $C_i^{X=\bar\eta^{\phi}}$ and   $C_i^{X=\bar\eta^{r}}$ are different  (and related among them because of the equations \eqref{sys_phi_r}).
Let us note that ${\mathcal A}$ can vanish (i.e., change its sign) at a certain value of $r_0$;  similarly, 
${\mathcal D}$ exists for $r_0>\frac{3}{2}r_s$ and vanishes at
\beq
r_0=r_*=\frac{5+\sqrt{2}}{4}r_s\approx 1.6036 r_s\,.
\eeq

\begin{figure}
    \centering
    \includegraphics[width=0.9\linewidth]{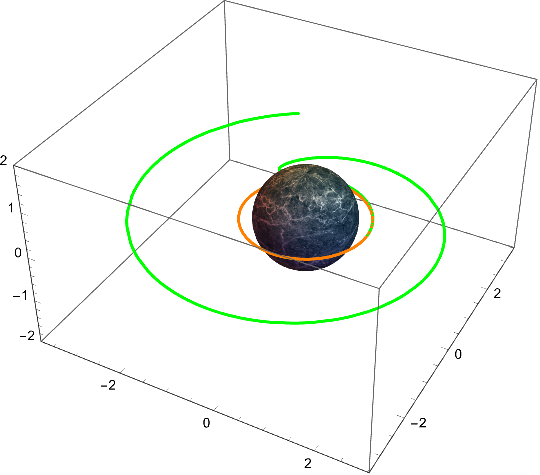}   
    \caption{Spin Deviation from the circular geodesic chosen parameters $r_b=1$, $r_s=0.8$, $r_0=1.25$. }
    \label{spindev}
\end{figure}

In order to have both the circular geodesic and its spin deviation starting from the same radial distance from the center at the same angle and with the same tangent vectors, we fix the following four boundary conditions
\beq\label{bcspin}
\bar\eta^r(0)=-\frac{\bar{B}}{\Sigma_r^2}\,,\quad \bar\eta^\phi(0)=0\,,\quad \frac{d\bar\eta^{r,\phi}(\tau)}{d\tau}\bigg|_{\tau=0}=0\,.
\eeq
This implies that two constants in the solution \eqref{gensol} will remain anyway arbitrary. 
Let us consider then the case $C_1=0=C_2$, i.e., the reduced equation $D_-(D_+( X(\tau)))=0$.
The solutions of the system \eqref{eqspin2} with the  boundary conditions \eqref{bcspin} (and, for example, in the region $\frac32 r_s <r_0<\frac{5+\sqrt{2}}{4}r_s$, see below) are
\bea
\bar \eta^{r}(\tau)
&=&
{\mathcal A}_r\Big[ \cosh\Omega_-\tau+\frac{2\Omega_+^2+7\Sigma_\phi^2}{2(\Omega_+^2-\Sigma_r^2)}\cosh\Omega_+\tau\Big]\,,
\nonumber\\
\bar \eta^{\phi}(\tau)&=&\mathcal{A}_\phi\Big[\sinh\Omega_-\tau-\frac{\Omega_-}{\Omega_+}\sinh\Omega_+\tau\Big]\,,
\eea
where 
\bea
{\mathcal A}_r &=& -\frac{2\bar{B}(\Omega_+^2-\Sigma_r^2)}{\Sigma_r^2\sqrt{\mathcal{D}}}\,, \nonumber\\
{\mathcal A}_\phi &=& \frac{9\bar{B}\Sigma_\phi^2}{B_2\Omega_-\sqrt{\mathcal{D}}}\,.
\eea
For completeness, one should distinguish the following three cases 
\bea
r_0&>&r_*,\quad \mathcal{D}<0,\quad\Omega_{\pm}\in \mathbb{C}\,,\nonumber\\
r_0&=&r_*,\quad \mathcal{D}=0,\quad\Omega_+{=}\Omega_-{=}\left(1{+}\frac{3}{\sqrt{2}}\right)^{1/2}\Sigma_\phi\,,\nonumber\\
r_0&<&r_*,\quad \mathcal{D}>0,\quad \Omega_{\pm}\in \mathbb{R}\,.
\eea
In the first case ($r_0>r_*$) both the radial and angular are oscillating functions with an (increasing) exponential damping in time, while in the last case ($r_0<r_*$) the behavior in $\tau$ is purely exponential, as it is shown in the plot in Fig. \ref{spindev}. The second (critical case, $r_0=r_*$, $\Omega_+=\Omega_-=\Omega_c =\frac{\sqrt{\mathcal A}}{2}$) requires a special treatment. Imposing the same boundary conditions as in \eqref{bcspin}, the solution of the system \eqref{sys_phi_r} can be written in the form
\bea
\bar{\eta}^r(\tau)&=&\mathcal{A}^r_c[4\cosh (\Omega_c\tau)+3\sqrt{2} \Omega_c\tau \sinh (\Omega_c\tau)]\,, \nonumber\\
\bar \eta^{\phi}(\tau)&=&\mathcal{A}^\phi_c [ \sinh (\Omega_c\tau)- \Omega_c\tau \cosh (\Omega_c\tau)]\,,
\eea
where
\beq
[\mathcal{A}^r_c,\mathcal{A}^\phi_c]=\frac{\bar{B}}{4\Sigma_r^2}\left[-1,\frac{9\Omega_c}{B_2}  \right]\,,
\eeq
with $B_2,\bar{B},\Sigma_r$ evaluated at $r=r_c$.

\begin{figure*}
\[
\begin{array}{cc}
    \includegraphics[width=0.3\linewidth]{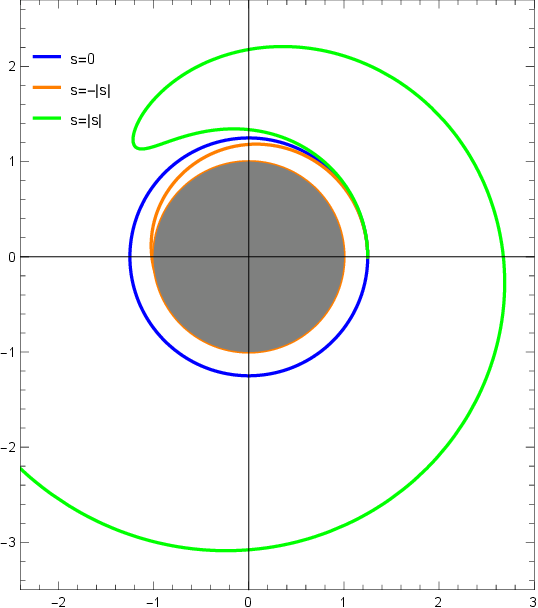}  
&
    \includegraphics[width=0.33\linewidth]{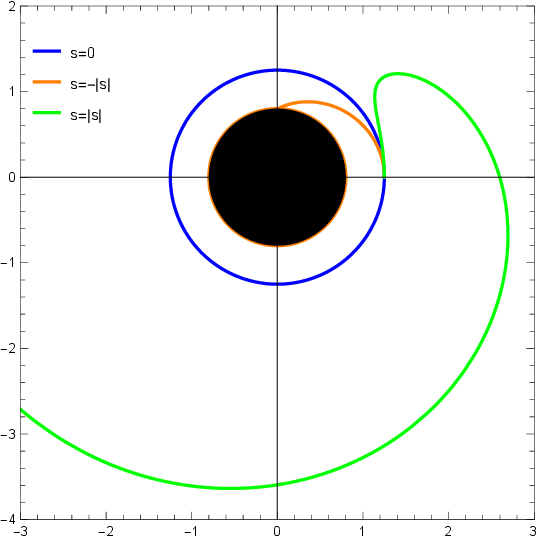}\cr  
(a) & (b) \cr
\end{array}
\]
\caption{\label{samepoint} Comparison of the geodesic deviations of three particles with different spin starting from the same initial point $r(0)=r_0$ and $\phi(0)=0$. The blue trajectories correspond to the particle moving on the circular geodesic while the green and orange trajectories refer to the particle with spin aligned in the same and opposite z direction respectively. The left panel (a) concerns motion in  a TS spacetime with parameters $r_b=1$, $r_s=0.8$, $r_0=1.25$ while the right panel (b) refers to Schwarzshild BH with the same parameters as the TS but now $r_b=0$.  }
\end{figure*}

\begin{figure*}
    \centering
    \includegraphics[width=0.22\linewidth]{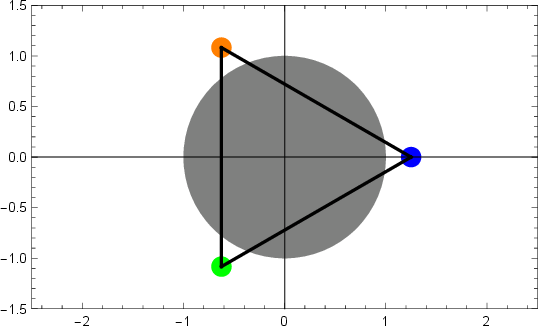}  
    \includegraphics[width=0.22\linewidth]{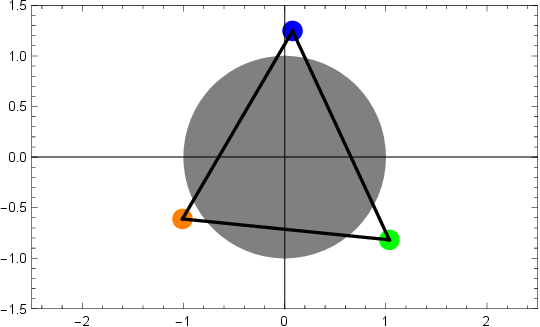}  
    \includegraphics[width=0.22\linewidth]{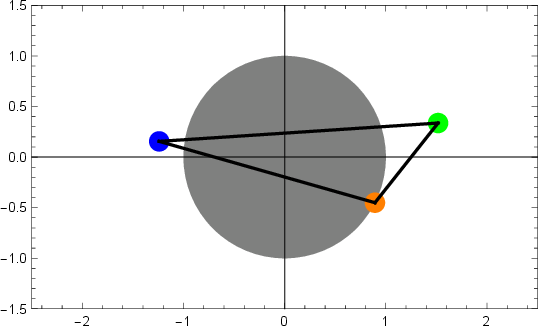}  
    \includegraphics[width=0.22\linewidth]{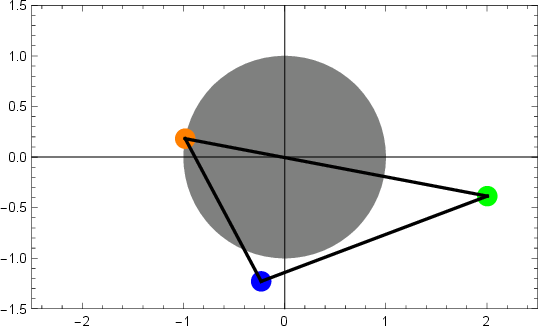}  
    \includegraphics[width=0.22\linewidth]{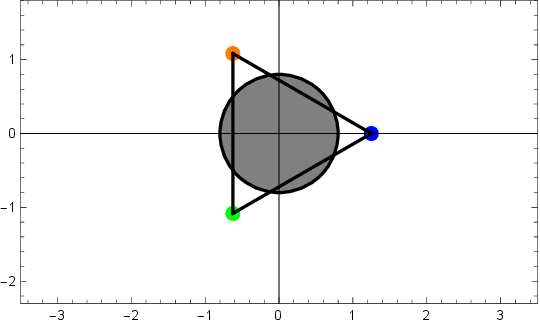}  
    \includegraphics[width=0.22\linewidth]{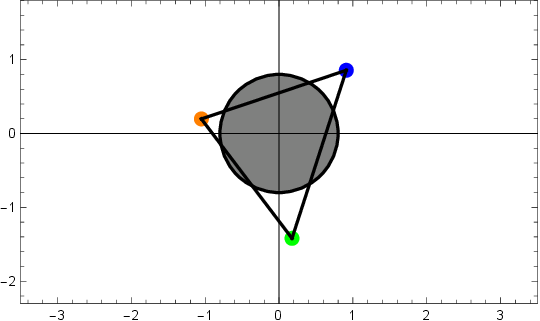}  
    \includegraphics[width=0.22\linewidth]{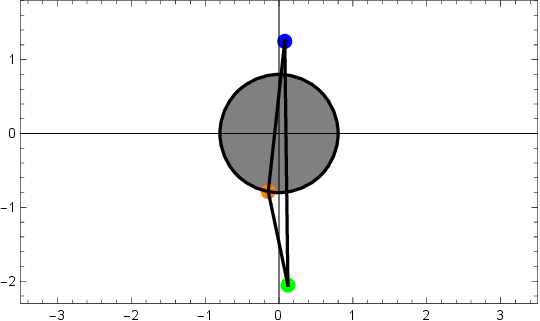}  
    \includegraphics[width=0.22\linewidth]{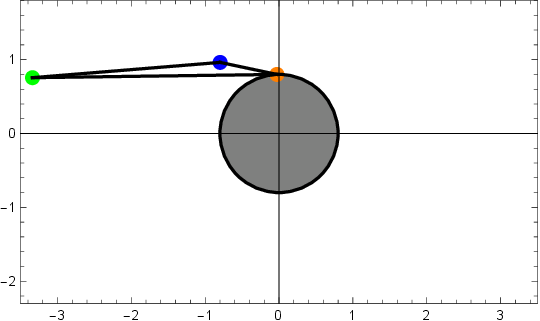}  
    \caption{ \label{snap} Evolution in time of three particles with different spin: blue dot represent the particle with zero spin, green (orange) dot the particle with spin with the same (opposite) direction as the z axis. The upper panel refer to a TS with parameters $r_b=1$, $r_s=0.8$, $r_0=1.25$ while the right panel refers to Schwarzshild BH with the same parameters as the TS but with $r_b=0$. The boundary conditions are chosen such that the initial radial distance is always $r_0$, while the initial angle are $\phi=0,2\pi/3,4\pi/3$.}
\end{figure*}

Finally, let us analyze the motion of particles within this family which have  different spin $s$. In particular, we consider  the geodesic ($s=0$) and two other general solutions having  spins  $s=\pm 1$ (in dimensionless units)  starting at the same radius, see Figs. \ref{samepoint}. The particle with positive spin deviates outward,  while the one with negative spin deviates inward and approaches $r=r_b$. In the analogous situation in the Schwarzschild BH case the particle with negative spin deviates inward and approaches the horizon $r=r_s$. The difference is the way in which this limiting radius is approached: In the TS case the orbit get closer and closer to $r_b$ tangentially while in the Schwarzschild case the approach at the horizon with a certain inclination. Indeed, in the TS spacetime the particle cannot enter the region $r\le r_b$ while in the Schwarzschild case the particle falls into the hole. 

We can consider the motion of these three particles as if they were initially at the vertices of an (ideal) equilateral triangle, inspired by a LISA experiment situation. The edges of the triangles are represented by ideal lines (solid), for graphical purposes only. The motion of the particles as described above implies variations to the edges of the triangle plus motions, corresponding to a continuous   squeezing and dilation during the motion. The situation is  illustrated in Fig. \ref{snap} where several snapshots of the motion are taken.

\section{Particles with an electromagnetic structure}\label{magncharge}

Let us study the motion of charged particles accelerated by the background electromagnetic field $F_{\alpha\beta}$
\beq\label{fma}
m a(U)^\alpha =q F^{\alpha}{}_{\beta}U^\beta\,, \quad F_{23}=-F_{32}=P\sin(\theta)\,,
\eeq
that is, explicitly
\bea\label{eomcharged}
\frac{d u^t(\tau)}{d\tau}&+&\frac{r_s u^r(\tau) u^t(\tau)}{r(\tau)(r(\tau)-r_s)}=0\,,\nonumber\\
\frac{d u^r(\tau)}{d\tau}&+&\frac{(r(\tau)-r_b)(r(\tau)-r_s)r_s (u^t(\tau))^2}{2r(\tau)^4}\nonumber\\
&-&\frac{(r(\tau)(r_b+r_s)-2r_b r_s)(u_r(\tau))^2}{2r(\tau)(r(\tau)-r_b)(r(\tau)-r_s)}\nonumber\\
&-&\frac{(r(\tau)-r_b)(r(\tau)-r_s)(u^\theta(\tau))^2}{r(\tau)}\nonumber\\
&-&\frac{(r(\tau){-}r_b)(r(\tau){-}r_s)\sin^2(\theta(\tau))(u^\phi(\tau))^2}{r(\tau)}{=}0\,,\nonumber\\
\frac{d u^\theta(\tau)}{d\tau}&+&\frac{2 u^r(\tau)u^\theta(\tau)}{r(\tau)}-\sin(\theta(\tau))\cos(\theta(\tau))(u^\phi(\tau))^2\nonumber\\
&=&\frac{\hat{q} P \sin(\theta(\tau))u^\phi(\tau)}{r(\tau)^2}\,,
\eea
and
\bea
\frac{d u^\phi(\tau)}{d\tau}&+&2\cot(\theta(\tau))u^\theta(\tau)u^\phi(\tau)+\frac{2u^r(\tau)u^\phi(\tau)}{r(\tau)}\nonumber\\
&=&-\frac{\hat{q} P u^\theta(\tau)}{r(\tau)^2 \sin(\theta(\tau))}\,,
\eea
where we introduced the (dimensionless) charge of the probe $\hat{q}=\frac{q}{m}$. The non-vanishing components of the acceleration are 
\bea
a(U)^\theta &=& \frac{\hat{q} P \sin(\theta)u^\phi(\tau)}{r^2}\,,\nonumber\\
a(U)^\phi &=&-\frac{\hat{q} P u^\theta(\tau)}{r^2 \sin(\theta)}\,.
\eea
Eqs. \eqref{eomcharged}  can be obtained from the Lagrangian
\bea
\mathcal{L}&=&\frac{1}{2}\Big[-f_s(r)(u^t(\tau))^2+\frac{(u^r(\tau))^2}{f_s(r)f_b(r)}+r^2(u^\theta(\tau))^2\nonumber\\
&+&r^2\sin(\theta)^2(u^\phi(\tau))^2\Big]-\hat{q} P\cos(\theta)u^\phi(\tau)\,,
\eea
with conjugate (canonical) momenta 
\bea
P_t&=&-E=-f_s(r) u^t(\tau)\,,\nonumber\\
P_r&=&\frac{u^r(\tau)}{f_s(r)f_b(r)}\,,\nonumber\\
P_\theta&=&r^2 u^\theta(\tau)\,,\nonumber\\
P_\phi&=&L=r^2\sin(\theta)^2 u^\phi(\tau)-\hat{q} P\cos(\theta) \,,
\eea
where $P_t=-E$ and $P_\phi=L$ are  conserved quantities as in the case of neutral particles (following geodesics).
The mass-shell constraint for timelike orbits, $2\mathcal{L}=-m^2$, allows us to re-express  the radial equation  \eqref{eomcharged} as

\begin{widetext}
\bea\label{systcharged}
\frac{dr(T)}{dT}&=&\pm \frac{\sqrt{(r{-}r_s)(r{-}r_b)}\sqrt{b^2(r_s-r)\left(\frac{\hat L^2}{s_\theta^{2}}(r-r_s)-r^2(r(\hat E^2-1)+r_s)\right)-v^2\hat E^2r^6\left(\frac{d\theta}{dT}\right)^2+b^2\hat P^2\hat{q}^2(r-r_s)^2\frac{c_\theta^2}{s_\theta^{2}}}}{v \hat E r^3}\,,\nonumber\\
\frac{d^2\theta(T)}{dT^2}&+&\frac{(3r_s-2r)}{(r_s-r)r}\frac{dr(T)}{dT}\frac{d\theta(T)}{dT}-\frac{b^2(r-r_s)^2(2(\hat L^2+\hat{q}^2\hat P^2)c_\theta+\hat{q} \hat P \hat L(3+c_\theta^2-s_\theta^2))}{2v^2\hat E^2r^6 s_\theta^3}=0\,,\nonumber\\
\frac{d\phi(T)}{dT}&=&\frac{b(r-r_s)(\hat L+\hat{q} \hat P c_\theta)}{v \hat E r^3s_\theta^2}\,,
\eea
\end{widetext}
where $\hat E=E/m$, $\hat L=L/m$, $\hat P=P/m$ and  we have introduced the compact notation $[s_\theta, c_\theta]=[\sin(\theta),\cos(\theta)]$ as well as the dimensionless coordinate $T=\frac{v t}{b}$. The $\pm$ in the radial equation correspond to integrating radially ingoing ($-$) or radially outgoing ($+$). We will mainly study the latter situation.

\begin{widetext}

\begin{figure*}
    \centering
    \includegraphics[width=0.22\linewidth]{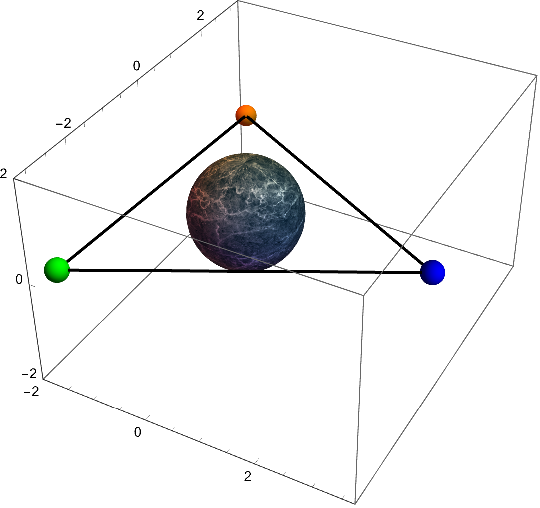}  
    \includegraphics[width=0.22\linewidth]{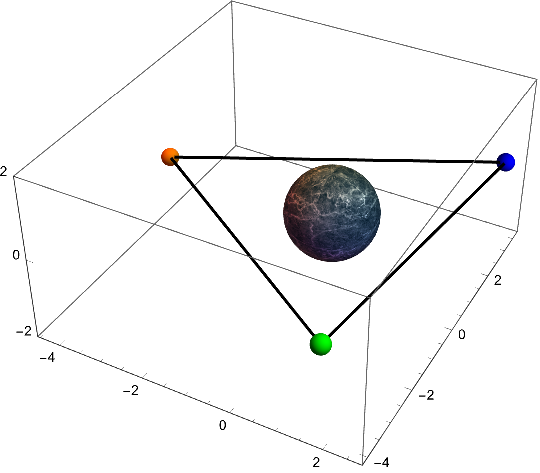}  
    \includegraphics[width=0.22\linewidth]{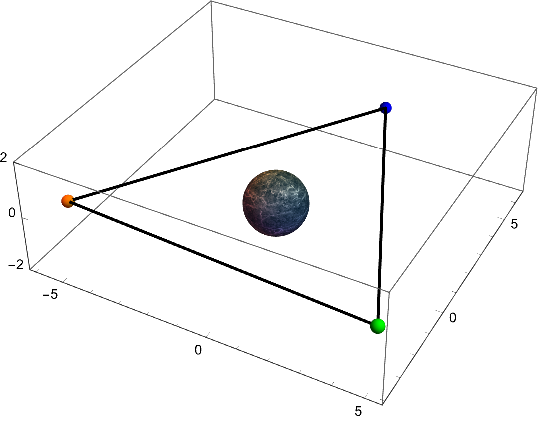}  
    \includegraphics[width=0.22\linewidth]{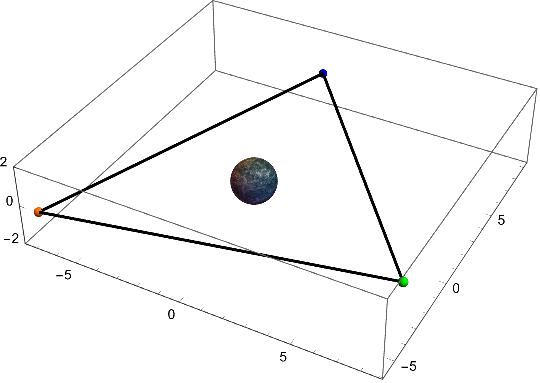}  
    \caption{ \label{accgeoplot2} Evolution in time of three particles with different magnetic charges: blue dot represents the neutral particle while orange (green) dot the particle with charge $q=(-)0.3$. The plots refer to a TS with parameters $v=0.5$, $b=5$, $\alpha=5/4$, $r_s=0.8$, $P=0.5$. The boundary conditions are chosen such that the initial angles are $\phi=0,2\pi/3,4\pi/3$.}
\end{figure*}

\end{widetext}

The coupled differential equations appearing in \eqref{systcharged} can be decoupled if we expand in small probe charge $\hat{q}$
\bea
r(T)&{=}&r_0(T){+}\hat{q}\,r_1(T){+}\hat{q}^2\,r_2(T){+}\mathcal{O}\left(\hat{q}^3\right)\,,\nonumber\\
\theta(T)&{=}&\frac{\pi}{2}{+}\hat{q}\,\theta_1(T){+}\hat{q}^2 \,\theta_2(T){+}\mathcal{O}\left(\hat{q}^3\right)\,,\nonumber\\
\phi(T)&{=}&\phi_0(T){+}\hat{q}\,\phi_1(T){+}\hat{q}^2\, \phi_2(T){+}\mathcal{O}\left(\hat{q}^3\right)\,.
\eea
Furthermore, each order in $\hat{q}$ expansion will be PM-expanded too
\bea
r_i(T)&=&r_{i,0}+\epsilon r_{i,1}+\epsilon^2 r_{i,2}+\mathcal{O}\left(\epsilon^3\right)\,,\nonumber\\
\theta_i(T)&=&\theta_{i,0}+\epsilon \theta_{i,1}+\epsilon^2 \theta_{i,2}+\mathcal{O}\left(\epsilon^3\right)\,,\nonumber\\
\phi_i(T)&=&\phi_{i,0}+\epsilon \phi_{i,1}+\epsilon^2 \phi_{i,2}+\mathcal{O}\left(\epsilon^3\right)\,.
\eea
The coefficients $r_{0,i}$ and $\phi_{0,i}$ for $i=0,1,2$ are displayed in Table \ref{tab:ri-phii-uti}. The $\theta$-equation simplifies as follows 
\be
\theta(T)=\frac{\pi}{2}+\hat{q}\frac{P\sqrt{1-v^2}}{b v}+\mathcal{O}\left(\epsilon^3, q^3\right)\,,
\ee
while $r_{1,i}=\phi_{1,i}=0$ and the other coefficients  $r_{2,j}$ and $\phi_{2,j}$ with $j=0,1,2$ are collected in Table \ref{rijphiij}. These accelerated orbits are represented in Fig.  \ref{accgeoplot2}.

\begin{table*}  
\caption{\label{rijphiij} List of the various coefficients $r_{i,j}$ and $\phi_{i,j}$ relative to charged orbits.}
\begin{ruledtabular}
\begin{tabular}{ll}
$\frac{r_{2,0}}{b}$ & $\frac{P^2(1-v^2)}{2b v^2}f_{\frac{1}{2}}$\\
$\frac{r_{2,1}}{b}$ & $\frac{P^2(1-v^2)(-1+v^2(\alpha+3))}{2 b v^2}(1-f_1+\tilde{f}^{\rm as1}_{\frac{3}{2}})$\\
$\frac{r_{2,2}}{b}$ & $\frac{P^2(1-v^2)(4v^2-1)}{4bv^2}f_{\frac{1}{2}}+\frac{P^2(1-v^2)(4v^2(\alpha+2)+v^4(\alpha^2-6\alpha-3)-1)}{4b v^2}\left(f_{\frac{1}{2}}-f_{\frac{3}{2}}\right)+\frac{3P^2v^2(1-v^2)(\alpha^2+2\alpha+5)}{4b}\left(\tilde{f}^{\rm at1}_{\frac{1}{2}}+\tilde{f}^{\rm at1}_{\frac{3}{2}}\right)$\\
   $$ & $+\frac{3P^2(v^2-1)(v^2(3+\alpha)-1)^2}{2bv^2}\tilde{f}^{\rm as1}_2+\frac{P^2(1-v^2)(v^2(\alpha+3)-1)^2}{4b v^2}\left(2 f^{\rm as2}_{\frac{3}{2}}-3f^{\rm as2}_{\frac{5}{2}}\right)$\\
 \hline
$\phi_{2,0}$ & $-\frac{P^2(1-v^2)}{2b^2 v^2}\left(\tilde{f}_1+f^{\rm at0}-\frac{\pi}{2}\right)$\\
$\phi_{2,1}$ & $-\frac{P^2(1-v^2)(2+\alpha v^2)}{b^2v^2}\tilde{f}_{\frac{3}{2}}-\frac{P^2(1-v^2)(1+v^2(\alpha+1))}{b^2v^2}\left(\tilde{f}_{\frac{1}{2}}-\tilde{f}_{\frac{3}{2}}\right)-\frac{P^2(1-v^2)(1-v^2(\alpha+3))}{b^2v^2}f^{\rm as1}_2$\\
$\phi_{2,2}$ & $-\frac{P^2(1-v^2)^2}{b^2v^2}\tilde{f}_2-\frac{3P^2(1-v^2)(4(3+\alpha)+v^2(3\alpha^2+2\alpha+3))}{8b^2}\tilde{f}_1-\frac{3P^2(1-v^2)(4(\alpha+3)+v^2(3\alpha^2+2\alpha+3))}{8b^2}\left(f^{\rm at0}-\frac{\pi}{2}\right)$\\
$$ & $+\frac{3P^2v^2(1-v^2)(\alpha^2+2\alpha+5)}{4b^2}\left(f^{\rm at1}_1+2f^{\rm at1}_2\right)+\frac{3P^2(1-v^2)^2(v^2(3+\alpha)-1)}{b^2v^2}f^{\rm as1}_{\frac{5}{2}}+\frac{P^2(1-v^2)(v^2(\alpha+3)-1)^2}{b^2v^2}\left(f^{\rm as1}_{\frac{3}{2}}-f^{\rm as1}_{\frac{5}{2}}\right)$\\
$$  &  $+\frac{2P^2(1-v^2)(v^2(\alpha+3)-1)^2}{b^2v^2}\tilde{f}^{\rm as2}_3$
\end{tabular}
\end{ruledtabular}
\end{table*}

%%%%%%%%%%%%%%%%%%%%%
\section{Electrically charged stringy probes in D=5}
\label{string}

The complete solution of the Einstein-Maxwell equations in $D=5$ admits both magnetic (2-form) and electric (3-form) fluxes \cite{Bah:2020pdz}. Therefore, as in Section \ref{magncharge}, and closely following the interpretation provided in \cite{Cipriani:2024ygw}, we can study deviations from geodesic motion for electrically charged probes.

However, the objects that naturally are coupled with a 2-form field  are one-dimensional objects, i.e., strings. In other words, the source \eqref{eqsource} for the electric field strenght \eqref{elfieldstr} is a charged line wound along the compact extra-direction $y$. In this description the quantity $Q$, which parametrizes the charge of the source, can be interpreted as a linear charge density of an infinitely extended wire.
Let us focus on rigid probes which propagate without oscillations. Under this assumption the embedding coordinate of their worldsheet in the TS geometry are 
\begin{equation}\label{embedding}
    X^M(\tau,\sigma)=(X^\mu(\tau),X^4=y+n\sigma R_y)\,,\quad \sigma\in[0,2\pi]\,,
\end{equation}
where the spacetime index is split the as  $M=(\mu,y)$, with $\mu=(t,r,\theta,\phi)$. Here $(\tau,\sigma)$ are the dimesionless worldsheet coordinates of the string and n is an integer corresponding to the winding number. 
The dynamics of a string in the TS background governed by the nonlinear sigma model
\bea
 S&=&\frac{1}{4\pi\alpha'}\int_\Sigma d\sigma d\tau\sqrt{-h}\Big[g_{MN}(X)\partial_a X^M\partial_b X^N h^{ab}\nonumber\\
 &+&A^{\rm (e)}_{MN}(X)\partial_a X^M\partial_b X^N \epsilon^{ab}\Big]\nonumber\\
&\equiv & S_G+S_B \,,
\eea
where 
\bea
S_G&=&  \int_\Sigma D_{(\sigma,\tau)}\, g_{MN}(X)\partial_a X^M\partial_b X^N h^{ab}\,, \nonumber\\
S_B&=&  \int_\Sigma D_{(\sigma,\tau)} \, A^{\rm (e)}_{MN}(X)\partial_a X^M\partial_b X^N \epsilon^{ab}\,,
\eea
with
\beq
D_{(\sigma,\tau)}= \frac{\sqrt{-h}}{4\pi\alpha'} d\sigma d\tau\,.
\eeq

Here
$\Sigma$ represents the world sheet of the string, $h^{ab}={\rm diag}[1,-1]$ (so that $h={\rm det}[h_{ab}]=-1$)
 is the world sheet metric, $\epsilon_{ab}$ is the 2-dimensional alternating symbol (defined such that $\epsilon^{\tau\sigma}=1$) and $\alpha'$ is the inverse string tension \cite{Callan:1985ia}. 
Moreover, $g_{MN}$ and $A^{\rm (e)}_{MN}$ are the metric and the gauge potential of the background, respectively. The effect of the gauge field on the conjugate momenta are $P_M=\partial L/\partial \dot{X}^M$, where $\dot{X}^M=\partial_\tau X^M$ can be studied by introducing the Lagrangian as
\begin{equation}
    S=\int d\tau \mathcal{L}\,,  
\end{equation}
with
\beq
\mathcal{L}=\mathcal{L}_G+\mathcal{L}_B\,,
\eeq
corresponding to $S=S_G+S_B$.
Setting $\alpha'=1$, the full Lagrangian in presence of the gauge potential \eqref{eqsource} and the embedding coordinate \eqref{embedding} appears to be
\bea\label{lbtemr}
\mathcal{L}_G&{=}&\frac{1}{2}g_{MN}(X)\partial_a X^M\partial_b X^N h^{ab}\nonumber\\
&=&\frac{1}{2}\Big[{-}f_s(r)\dot{t}^2{+}\frac{\dot{r}^2}{f_s(r)f_b(r)}{+}r^2(\dot{\theta}^2{+}\sin^2\theta\dot{\phi}^2)\nonumber\\
&{-}&f_b(r)q^2\Big]\,,\nonumber\\
\mathcal{L}_B&=&\frac{1}{2}A^{\rm (e)}_{MN}(X)\partial_a X^M\partial_b X^N \epsilon^{ab}=-\frac{qQ}{r}\dot{t}\,,
\eea
where we defined the winding charge of the stringy probe as
\begin{equation}
    q=n R_y\,.
\end{equation}
Introducing the canonical conjugate momenta
\be
P_\mu=\frac{\partial\mathcal{L}}{\partial\dot{x}^\mu}\,,
\ee
and consequently the conserved energy ($P_t=-E$) and angular momentum ($P_\phi=L$) the dynamics of the center-of-mass of an electrically charged string is described by the following set of equations
\bea\label{dotxel}
\dot{t}=\frac{E-\frac{q Q}{r}}{f_s(r)}\,&,&\quad \dot{r}=f_s(r)f_b(r) P_r\,,\nonumber\\
\dot{\theta}=\frac{P_\theta}{r^2}\,&,&\quad \dot{\phi}=\frac{L}{r^2\sin^2\theta}\,.
\eea
So the additional contribution due to the electric field source \eqref{lbtemr} depends solely on $\dot{t}$, all the conjugate momenta previously defined in the absence of ${\mathcal L}_B$ remain unchanged by the coupling with the gauge field, except for the conjugate momentum in the $t$ direction. So it is possible to account for the presence of the additional contribution (i.e. ${\mathcal L}_B$) by making the following replacement in the canonical momentum along $t$ direction
\begin{equation}
    P_t\rightarrow P_t+\frac{qQ}{r}\,.
\end{equation}
This can be interpreted from a wave perspective as the substitution of the ordinary derivative with the covariant derivative, in order to account for the minimal coupling of the test field with the source.

The extra non derivative term $q^2 f_b(r)$ appearing in $\mathcal{L}_G$ can be though as a contribution to the mass due to the wrapping of the stringy probe around the compact $y$ direction.

In general, the string mass in light-cone gauge quantization is determined by the contributions of string oscillators \cite{Polchinski:1998rq}, which, within our effective approach, can be interpreted as the intrinsic mass of the stringy probe. Additionally, the presence of a compact direction ({\rm i.e.},
$y$) gives rise to extra mass contributions associated with winding modes and Kaluza-Klein momentum along this direction. These contributions are inherently related through a T-duality transformation \cite{Becker:2006dvp}.

\subsection{Circular orbits and Lyapunov exponent}

From Legendre transformation of the Lagrangian \eqref{lbtemr}, the Hamiltonian reads
\bea
\mathcal{H}&=&\frac{1}{2}\Bigg[-\frac{\left(E-\frac{q Q}{r}\right)^2}{f_s(r)}+f_s(r)f_b(r) P_r^2+\frac{P_\theta^2}{r^2}\nonumber\\
&+&\frac{L^2}{r^2\sin^2\theta}+f_b(r)q^2\Bigg]\,,
\eea
so that the Hamiltonian mass shell condition $2\mathcal{H}=-m^2$
is separated as follows
\bea
P_r^2&=&Q_R(r)=\frac{ (rE-qQ)^2}{r^2 (f_s(r))^2 f_b(r)}-\frac{K^2}{r^2 f_s(r)f_b(r)}\nonumber\\
&-&\frac{m^2  }{ f_s(r)f_b(r)}-\frac{q^2  }{ f_s(r) }\,,\nonumber\\
P_\theta^2&=&K^2-\frac{L^2}{\sin^2\theta}\,.
\eea
where $m$ can be interpreted as the intrinsic mass of the string. Again the spherical symmetry allows us to limit considerations to equatorial orbits ($\theta=\pi/2$) with $K=L$. A double zero of the radial potential  $Q_R$ signals the existence of a (unstable) circular geodesic,
\begin{equation}
    Q_R(r_c,L_c,E_c)=Q_R'(r_c,L_c,E_c)=0\,.
\end{equation}
Let us schematically decompose $Q_R$ as
\bea
Q_R(r)&=&\frac{(E^2-m^2-q^2)}{r^3 (f_s(r))^2 f_b(r)}P_3(r)\,,\nonumber\\
P_3(r)&=&r^3+A r^2+Br+C\,,\nonumber\\
A&=&\frac{q^2 \left(r_b+r_s\right)+m^2 r_s-2 E q Q}{E^2-m^2-q^2}\,,\nonumber\\
B&=&-\frac{q^2 \left(r_b r_s-Q^2\right)+L^2}{E^2-m^2-q^2}\,,\nonumber\\
C&=&\frac{L^2 r_s}{E^2-m^2-q^2}\,.
\eea
The polynomial $P_3(r)$  is conveniently studied by   introducing the coordinate 
\begin{equation}
r=t-\frac{A}{3}\,,
\end{equation}
needed to recast the cubic in its \lq\lq depressed form"
\bea
P_3(t)&=&t^3+p t+q\nonumber\\
p&=&B-\frac{A^2}{3}\qquad q=\frac{2A^3}{27}-\frac{AB}{3}+C\,,
\eea
with discriminant given by
\begin{equation}
\Delta=-4p^3-27 q^2\,.
\end{equation}
In the case $p\neq 0$ and $\Delta=0$, the cubic equation has a double root and can be written as follows
\begin{equation}\label{doubleroot}
P_3(t)=\left(t-\frac{3q}{p}\right)\left(t+\frac{3q}{2p}\right)^2\,.
\end{equation}
So from Eq. \eqref{doubleroot} we can identify the critical radius as the double zero of the cubic equation and an (quite implicit) expression for the critical angular momentum 
\bea
r_c&=&\frac{9C-AB}{2(A^2-3B)}\,,\nonumber\\
C&=&\frac{9AB-2A^3\pm2(A^2-3B)^{3/2}}{27}\,,
\eea
where the second equality comes from the condition $\Delta=0$. Denoting the simple root in \eqref{doubleroot} as
\begin{equation}
    r_3=\frac{4AB-9C-A^3}{A^2-3B}\,,
\end{equation}
 the polynomial $P_3(r)$ becomes
\begin{equation}
    P_3(r)=(r-r_c)^2(r-r_3)\,,
\end{equation}
where $r_3=-2r_c$ and
\bea
r_c&\approx& \frac{3r_s}{2}-q\frac{3\sqrt{3}Qr_s}{2J}+q^2\frac{r_s(32Q^2+3r_s(5r_b+9r_s))}{8J^2}\nonumber\\
&+&m^2\frac{27r_2^3}{8J^2}\,,\nonumber\\
E_c&\approx&\frac{2J}{3\sqrt{3}r_s}+q\frac{2Q}{3r_s}+m^2\frac{\sqrt{3}r_s}{4J}+\nonumber\\
&+&q^2\frac{4Q^2-6r_br_s+9r_s^2}{12\sqrt{3}J r_s}\,,
\eea
up to order $\mathcal{O}(m^2,q^2)$ included. The chaotic behavior in the  region close the unstable critical geodesic can be described in terms on the Lyapunov exponent
\begin{equation}
    \frac{dr}{dt}  
=-\frac{f_s(r)^2f_b(r)\sqrt{Q_R(r)}}{E-\frac{qQ}{r}}\bigg|_{r\to r_c}\sim-\lambda(r-r_c)\,,
\end{equation}
so that
\begin{equation}\label{lyapel}
\lambda=\frac{\sqrt{3}f_s(r_c)\sqrt{f_b(r_c)}\sqrt{E_c^2-m^2-q^2}}{(r_c E_c-qQ)}\,.
\end{equation}
In expanded form this is equivalent to
\bea
\lambda&\approx& \frac{2\zeta}{9r_s}+q\frac{Q(\zeta^2-3)}{3\sqrt{3}J r_s \zeta}+m^2\frac{r_s(3-2\zeta^2)}{4\zeta J^2}\nonumber\\
&{-}&\frac{3r_s^2\zeta^2(5\zeta^4{+}36\zeta^2{-}99){+}2Q^2(157\zeta^4{-}42\zeta^2{+}81)}{216 r_s J^2 \zeta^3}\,,\qquad
\eea
having defined
\beq
\zeta=\sqrt{\frac{3r_s-2r_b}{r_s}}\,.
\eeq
Note that the Lyapunov exponent \eqref{lyapel} in the limit $r_b=m=q=0$ and $r_s=2M$ reproduces the Lyapunov exponent at the Schwarzschild shadow $\lambda_{\rm Sch}=\frac{1}{3\sqrt{3}M}$ \cite{Heidmann:2022ehn}.
It is known that the Lyapunov exponent plays a role in the estimate of QNMs frequencies, whose  imaginary part describes the damping in time of the modes \cite{Bianchi:2021mft,Heidmann:2022ehn,Bianchi:2023sfs}. Actually, in this approximation, the QNMs frequencies have the form \cite{Cardoso:2008bp}
\begin{equation}
    \omega_{_{\rm WKB}}=E_c-i\left(n+\frac{1}{2}\right)\lambda\,,
\end{equation}
where $E_c$ is exactly the   critical energy corresponding to the photon sphere.
\subsection{Deviations}

\begin{figure}
    \centering
    \includegraphics[width=0.4\linewidth]{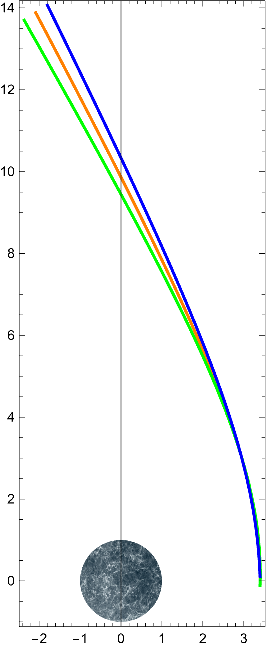}  
    \caption{\label{plotstringel1} Trajectories followed by the center of mass of the electrically charged string for $v=0.5$, $b=5$, $\alpha=5/4$, $r_s=0.8$, $Q=0.5$. The orange plot correspond to neutral string $q=0$, the green (blue) one to positively (negatively) charged string $q=(-)0.3$. The boundary conditions are such that $\phi(0)=0$ in all cases.}
\end{figure}

\begin{figure*}
    \centering
    \includegraphics[width=0.2\linewidth]{{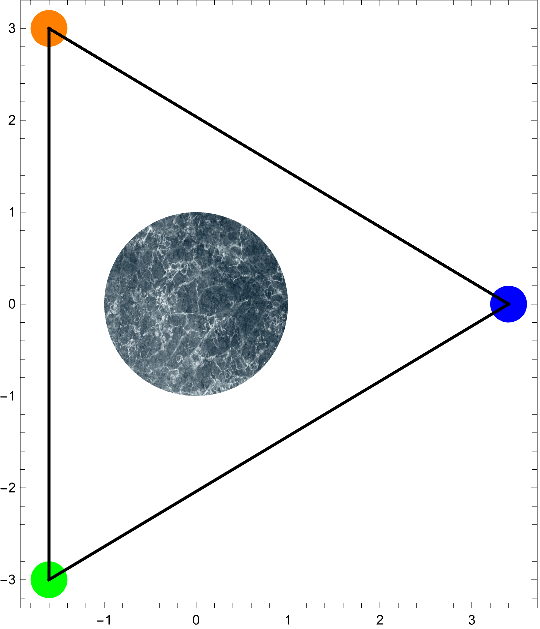}}  
    \includegraphics[width=0.23\linewidth]{{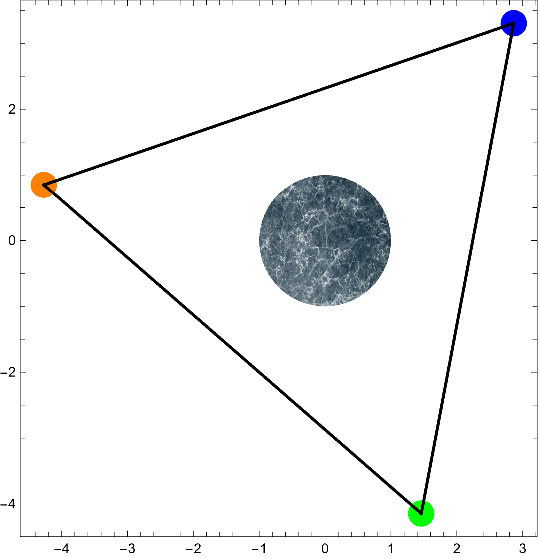}}  
    \includegraphics[width=0.24\linewidth]{{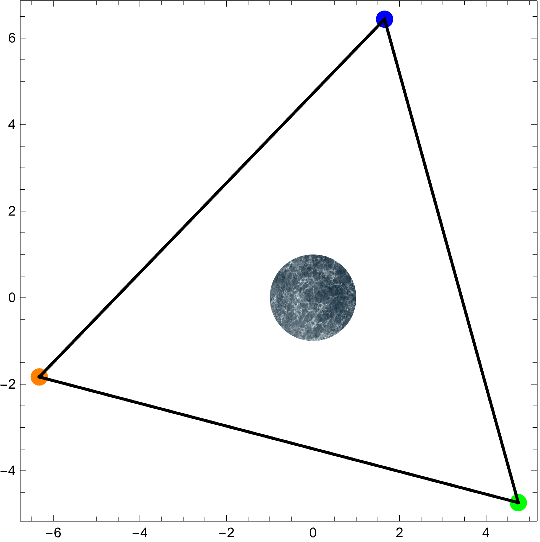}} 
    \includegraphics[width=0.27\linewidth]{{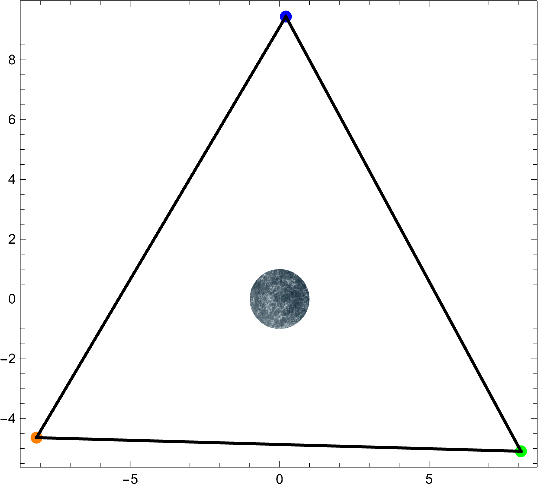}} 
    \caption{\label{plotstringel2}Evolution in time of the centers of mass of three strings with different electric charges: blue dot represents the neutral string while orange (green) dot the string with charge $q=(-)0.3$. The plots refer to a TS with parameters $v=0.5$, $b=5$, $\alpha=5/4$, $r_s=0.8$, $Q=0.5$. The boundary conditions are chosen such that the initial angles are $\phi=0,2\pi/3,4\pi/3$.}
\end{figure*}

The (equatorial) equations of motion  (with the mass-shell condition $2\mathcal{L}=-m^2$) can be written in terms of the dimensionless variable $T=vt/b$ as:
\begin{widetext}
    \bea\label{solstreq}
\frac{dr(T)}{dT}&=&-\frac{bf_s(r)\sqrt{f_b(r)}}{v \sqrt{r}(\hat{E} r-\hat{q} Q)}\sqrt{(\hat{E}^2-1-\hat{q}^2)r^3+(r_s+\hat{q}^2(r_b+r_s)-2Q\hat{q}\hat{E})r^2-(\hat{q}^2(Q^2-r_br_s)-\hat{L}^2)r+\hat{L}^2 r_s}\,,\nonumber\\
\frac{d\phi(T)}{dT}&=&\frac{b\hat{L}f_s(r)}{vr(\hat{E} r-\hat{q}Q)}\,,
    \eea
where we introduced the dimensionless variables $\hat{E}=E/m$   and $q=\hat{q}/m$ as well as the rescaled angular momentum $\hat{L}=L/m$ with the dimensions of a length. The solutions to these equations can be found perturbatively in a large-$b$ (PM-expansion) or equivalently small $\epsilon=\frac{r_s}{2b v^2}$, and for small charge of the stringy probe $q$. The solutions at order $q^0$ and in PM-expanded form are already collected in Table \ref{tab:ri-phii-uti}, while the deviations due to $q$ and at up to the first order in $\epsilon$ are displayed in Table \ref{tabstrel}. The plots of the trajectories followed by the center of masses of the electrically charged strings are represented in Fig. \ref{plotstringel1} and \ref{plotstringel2}.

\begin{table*}  
\caption{\label{tabstrel} Perturbative solutions the system \eqref{solstreq}. $r_{i,j}$ refer to the coefficients $q^i$ and $\epsilon^j$of the expanded solution.
}
\begin{ruledtabular}
\begin{tabular}{ll}
$\frac{r_{1,0}}{Q\sqrt{1-v^2}}$ & $ 1-\frac{\sqrt{1+T^2}{\rm arcsinh}(T)}{T}$\\
$\frac{r_{1,1}}{Q\sqrt{1-v^2}}$ & $\frac{v^2(3+\alpha)}{2\sqrt{1+T^2}}+\frac{(1-v^2(3+\alpha)){\rm arcsinh}^2(T)}{T^2\sqrt{1+T^2}}+\frac{\sqrt{1+T^2}(-2+v^2(3+\alpha)){\rm arctan}(T)}{2T} $\\
$\frac{b \phi_{1,0}}{Q\sqrt{1-v^2}}$ & $-\frac{T}{\sqrt{1+T^2}}-2{\rm arcsinh}(T)(\log(\sqrt{1+T^2})-\log[T]-i\pi)-{\rm Li}_2(-(T+\sqrt{1+T^2})^2)+{\rm Li}_2((T+\sqrt{1+T^2})^2)$\\
$\frac{\phi_{1,1}}{Q\sqrt{1-v^2}}$ & $-\frac{T \left((\alpha +1) v^2+1\right)}{T^2+1}+\frac{1}{12} i \pi ^2 \left((\alpha +3) v^2-2\right)+\frac{{\rm arcsinh}(T)^2 \left(2-2 (\alpha +3) v^2\right)}{T^3+T}-2\left((4 \alpha +5) v^2+3\right) {\rm arctan}\left(\frac{\sqrt{T^2+1}-1}{T}\right) +$\\
$$ & $\frac{1}{2} {\rm arctan}(T) \left(\frac{1}{2} \left(3 \log (1+i T)-4 \log (2 T)+\log
   (T+i)-\frac{5 i \pi }{2}\right) \left((\alpha +3) v^2-2\right)-\left((\alpha -1)
   v^2\right)-\alpha  v^2-5 v^2\right)+$\\
$$ & ${\rm arcsinh}(T) \left(\frac{T^2 \left((4 \alpha +6) v^2+2\right)+4 \left(T^2+1\right)^{3/2}
   \left(\log \left(\sqrt{T^2+1}+T-1\right)-\log \left(\sqrt{T^2+1}+T+1\right)\right)
   \left((\alpha +3) v^2-1\right)+(5 \alpha +9) v^2+1}{\left(T^2+1\right)^{3/2}}\right.$\\
$$ & $\left.-12
   \left(v^2-1\right) {\rm arctanh}\left(T-\sqrt{T^2+1}\right)\right)+2 \text{Li}_2\left(T-\sqrt{T^2+1}\right) \left((2 \alpha +3) v^2+1\right)$\\
$$ & $-2
   \text{Li}_2\left(\frac{1}{T+\sqrt{T^2+1}}\right) \left((2 \alpha +3)
   v^2+1\right)+\frac{1}{2} i \text{Li}_2\left(\frac{T+i}{i-T}\right) \left((\alpha +3)
   v^2-2\right)$
\end{tabular}
\end{ruledtabular}
\end{table*}
\end{widetext}

\section{Discussion}

Motivated by the lack of explicit analytic results concerning accelerated motions in a TS spacetime, we have analyzed the behaviour of spinning particles and electrically charged particles in such a spacetime. This study has shown how deviations  from geodesic motion can be relevant in order to characterize how TS differs from a Schwarzschild BH. In order to have  general information, we have framed our analysis in the context of relativistic \lq\lq deviations" solving the geodesic deviation equation.
In the latter case as well as for the motion of charged particles we have considered hyperbolic-like orbits treated in a large impact parameter expansion limit, i.e., within what is called Post-Minkowskian approximation limit. For spinning particles we have analyzed deviations from circular geodesic orbits (circular orbits are also allowed for spinning particles), a simpler but very interesting situation. We have defined a family of spinning particles with orbits parametrized by the spin itself (besides a proper time parametrization) and discussed what happens to a bunch of such particles starting at the same point when they have different spin. In particular we have examined the approach to the boundary (cap, spherical surface) of the spacetime in the TS case which turns out to be tangential-like. To better illustrate the deviations we have discussed the case of three particles evolving from a geometrically simple situation: a triangle (made of ideal edges, inspired by a LISA-like experimental situation), whose evolution corresponds to a continuous expansion and squeezing of the edges themselves.
As a general result we find that 1) while a BH captures particles, the TS only allows for passages which (smoothly) skim the cap, 2) deviations due to the TS structure correspond to small additional corrections to the Schwarzschild case, mostly interesting from a theoretical point of view. 
Finally, an appendix summarizes useful geometrical information for the $(t,r,y)$ part of the TS metric.

\section*{Acknowledgments}

We thank A. Geralico 
for useful comments. D. B.  acknowledges sponsorship of
the Italian Gruppo Nazionale per la Fisica Matematica
(GNFM) of the Istituto Nazionale di Alta Matematica
(INDAM).  

\appendix

\section{Geometrical properties of the 3d line element $ds_{(t,r,y)}^2$}
\label{dsquad_try}
We have defined 
\beq
\label{3-met2}
ds_{(t,r,y)}^2=-f_s(r)dt^2+\frac{dr^2}{f_s(r)f_b(r)}+f_b(r)dy^2\,,
\eeq
as the metric induced on the $\theta$=constant and $\phi$=constant hypersurfaces, with
\bea
{}^{(3)}g_{tt}&=&-f_s(r)\,,\quad {}^{(3)}g_{rr}= [f_s(r)f_b(r)]^{-1}\,,\nonumber\\ 
{}^{(3)}g_{yy}&=& f_b(r),
\eea
and ${\rm det}(g)=-1 $.
This metric is curved, with Ricci scalar given by
\beq
R=\frac{2( r_s+r_b)}{r^3}-\frac{11r_sr_b}{2r^4} \,,
\eeq
and to the curvature contribute either both the arithmetic and the geometric  averages of $r_b$ and $r_s$,
\beq
 r_{bs}=\frac12(r_b+r_s)\,,\qquad \bar r_{bs}=\sqrt{r_br_s}\,,
\eeq
namely
\beq
R=4 \frac{r_{bs}}{r^3}-\frac{11}{2} \frac{\bar r_{bs}^2 }{r^4} \,.
\eeq
The Cotton-York tensor $Y^{\alpha\beta}$, whose expression in $d=3$ uses the \lq\lq Symmetric curl" (or Scurl) operation,
\beq
Y^{\alpha\beta}=\eta^{\gamma \delta (\alpha}R^{\beta)}{}_{\gamma; \delta}\equiv -{\rm Scurl}[{\rm Ricci}]^{\alpha\beta}
\eeq
has a single nonvannishing component
\beq
Y^{0 y}=\frac{r_b}{2r^4}f_b(r) f_{11/6}(r)\,,
\eeq
implying that the metric \eqref{3-met2} is not conformally flat.

The timelike geodesics read
\bea
U=\frac{E}{f_s(r(\tau))}\partial_t +\frac{dr}{d\tau}\partial_r+\frac{P_y}{f_b(r(\tau))}\partial_y \,,
\eea
with
\beq
\left(\frac{dr}{d\tau}\right)^2=E^2f_b(r)-P_y^2f_s(r) -f_s(r)f_b(r)\,.
\eeq
Circular geodesics (along $y$) exist at $r(\tau)=r_0$,
\bea
U=\frac{E}{f_s(r_0)}\partial_t+\frac{P_y}{f_b(r_0)}\partial_y \,,
\eea
with corresponding completely covariant form (denoted by the symbol $\flat$) with constant components \cite{Bini:2012msy}
\beq
U^\flat =-E dt +P_y dy\,,
\eeq
so that 
\beq
dU^\flat =0\,.
\eeq
Because of the general relation (see, e.g., Eq. (7.1) of Ref. \cite{Jantzen:1992rg})
\beq
dU^\flat = U^\flat \wedge g(U) +{}^{*(U)}H(U)\,,
\eeq
where $g(U)=-\nabla_U U$ (gravitoelectric field, i.e., minus the acceleration of the $U$ congruence)  and $H(U)^\alpha=\frac12 \eta(U)^{\alpha\beta\gamma}[dU]_{\beta\gamma}$ with $\eta(U)^{\alpha\beta\gamma}=U_\sigma \eta^{\sigma\alpha\beta\gamma}$ (gravitomagnetic  field, i.e., the spatial dual of the vorticity of the $U$ congruence)  the fact that in our case $dU^\flat=0$ implies that
the $2$-parameter congruence of these orbits is  geodesic ($g(U)=0$) and  irrotational ($H(U)=0$).


\begin{thebibliography}{99}

 

%\cite{Bah:2020ogh}
\bibitem{Bah:2020ogh}
I.~Bah and P.~Heidmann,
``Topological Stars and Black Holes,''
Phys. Rev. Lett. \textbf{126}, no.15, 151101 (2021)
doi:10.1103/PhysRevLett.126.151101
[arXiv:2011.08851 [hep-th]].
%55 citations counted in INSPIRE as of 03 May 2025

%\cite{Bah:2020pdz}
\bibitem{Bah:2020pdz}
I.~Bah and P.~Heidmann,
``Topological stars, black holes and generalized charged Weyl solutions,''
JHEP \textbf{09}, 147 (2021)
doi:10.1007/JHEP09(2021)147
[arXiv:2012.13407 [hep-th]].
%44 citations counted in INSPIRE as of 12 May 2025

%\cite{Bah:2023ows}
\bibitem{Bah:2023ows}
I.~Bah and P.~Heidmann,
``Geometric resolution of the Schwarzschild horizon,''
Phys. Rev. D \textbf{109}, no.6, 066014 (2024)
doi:10.1103/PhysRevD.109.066014
[arXiv:2303.10186 [hep-th]].
%20 citations counted in INSPIRE as of 03 May 2025

%\cite{Heidmann:2022ehn}
\bibitem{Heidmann:2022ehn}
P.~Heidmann, I.~Bah and E.~Berti,
``Imaging topological solitons: The microstructure behind the shadow,''
Phys. Rev. D \textbf{107}, no.8, 084042 (2023)
doi:10.1103/PhysRevD.107.084042
[arXiv:2212.06837 [gr-qc]].
%21 citations counted in INSPIRE as of 03 May 2025

%\cite{Bianchi:2023sfs}
\bibitem{Bianchi:2023sfs}
M.~Bianchi, G.~Di Russo, A.~Grillo, J.~F.~Morales and G.~Sudano,
``On the stability and deformability of top stars,''
JHEP \textbf{12}, 121 (2023)
doi:10.1007/JHEP12(2023)121
[arXiv:2305.15105 [gr-qc]].
%30 citations counted in INSPIRE as of 23 Apr 2025

%\cite{Heidmann:2023ojf}
\bibitem{Heidmann:2023ojf}
P.~Heidmann, N.~Speeney, E.~Berti and I.~Bah,
``Cavity effect in the quasinormal mode spectrum of topological stars,''
Phys. Rev. D \textbf{108}, no.2, 024021 (2023)
doi:10.1103/PhysRevD.108.024021
[arXiv:2305.14412 [gr-qc]].
%21 citations counted in INSPIRE as of 12 May 2025

%\cite{DiRusso:2024hmd}
\bibitem{DiRusso:2024hmd}
G.~Di Russo, F.~Fucito and J.~F.~Morales,
%``Tidal resonances for fuzzballs,''
JHEP \textbf{04}, 149 (2024)
doi:10.1007/JHEP04(2024)149
[arXiv:2402.06621 [hep-th]].
%12 citations counted in INSPIRE as of 12 May 2025

%\cite{Cipriani:2024ygw}
\bibitem{Cipriani:2024ygw}
A.~Cipriani, C.~Di Benedetto, G.~Di Russo, A.~Grillo and G.~Sudano,
``Charge (in)stability and superradiance of Topological Stars,''
JHEP \textbf{07}, 143 (2024)
doi:10.1007/JHEP07(2024)143
[arXiv:2405.06566 [hep-th]].
%12 citations counted in INSPIRE as of 23 Apr 2025

%\cite{Bena:2024hoh}
\bibitem{Bena:2024hoh}
I.~Bena, G.~Di Russo, J.~F.~Morales and A.~Ruip\'erez,
``Non-spinning tops are stable,''
JHEP \textbf{10}, 071 (2024)
doi:10.1007/JHEP10(2024)071
[arXiv:2406.19330 [hep-th]].
%11 citations counted in INSPIRE as of 03 May 2025

%\cite{Bianchi:2024vmi}
\bibitem{Bianchi:2024vmi}
M.~Bianchi, D.~Bini and G.~Di Russo,
``Scalar perturbations of topological-star spacetimes,''
Phys. Rev. D \textbf{110}, no.8, 084077 (2024)
doi:10.1103/PhysRevD.110.084077
[arXiv:2407.10868 [gr-qc]].
%11 citations counted in INSPIRE as of 03 May 2025

%\cite{Bianchi:2024rod}
\bibitem{Bianchi:2024rod}
M.~Bianchi, D.~Bini and G.~Di Russo,
``Scalar waves in a topological star spacetime: Self-force and radiative losses,''
Phys. Rev. D \textbf{111}, no.4, 044017 (2025)
doi:10.1103/PhysRevD.111.044017
[arXiv:2411.19612 [gr-qc]].
%7 citations counted in INSPIRE as of 12 May 2025

%\cite{DiRusso:2025lip}
\bibitem{DiRusso:2025lip}
G.~Di Russo, M.~Bianchi and D.~Bini,
``Scalar waves from unbound orbits in a TS spacetime: PN reconstruction of the field and radiation losses in a self-force approach,''
[arXiv:2502.21040 [gr-qc]].
%2 citations counted in INSPIRE as of 03 May 2025

%\cite{Lunin:2001jy}
\bibitem{Lunin:2001jy}
O.~Lunin and S.~D.~Mathur,
``AdS / CFT duality and the black hole information paradox,''
Nucl. Phys. B \textbf{623}, 342-394 (2002)
doi:10.1016/S0550-3213(01)00620-4
[arXiv:hep-th/0109154 [hep-th]].
%574 citations counted in INSPIRE as of 12 May 2025

%\cite{Bena:2007kg}
\bibitem{Bena:2007kg}
I.~Bena and N.~P.~Warner,
``Black holes, black rings and their microstates,''
Lect. Notes Phys. \textbf{755}, 1-92 (2008)
doi:10.1007/978-3-540-79523-0\_1
[arXiv:hep-th/0701216 [hep-th]].
%429 citations counted in INSPIRE as of 12 May 2025

%\cite{Skenderis:2008qn}
\bibitem{Skenderis:2008qn}
K.~Skenderis and M.~Taylor,
``The fuzzball proposal for black holes,''
Phys. Rept. \textbf{467}, 117-171 (2008)
doi:10.1016/j.physrep.2008.08.001
[arXiv:0804.0552 [hep-th]].
%353 citations counted in INSPIRE as of 12 May 2025

%\cite{Bianchi:2022qph}
\bibitem{Bianchi:2022qph}
M.~Bianchi and G.~Di Russo,
``2-charge circular fuzz-balls and their perturbations,''
JHEP \textbf{08}, 217 (2023)
doi:10.1007/JHEP08(2023)217
[arXiv:2212.07504 [hep-th]].
%29 citations counted in INSPIRE as of 12 May 2025

%\cite{Chandrasekhar:1985kt}
\bibitem{Chandrasekhar:1985kt}
S.~Chandrasekhar,
``The mathematical theory of black holes''
(Oxford Classic Texts in the Physical Sciences)
Clarendon Press, 1998 - 646 pages
ISBN: 9780198503705
%363 citations counted in INSPIRE as of 12 May 2025

%\cite{Bianchi:2017sds}
\bibitem{Bianchi:2017sds}
M.~Bianchi, D.~Consoli and J.~F.~Morales,
%``Probing Fuzzballs with Particles, Waves and Strings,''
JHEP \textbf{06}, 157 (2018)
doi:10.1007/JHEP06(2018)157
[arXiv:1711.10287 [hep-th]].
%37 citations counted in INSPIRE as of 16 May 2025

%\cite{Bianchi:2021yqs}
\bibitem{Bianchi:2021yqs}
M.~Bianchi and G.~Di Russo,
``Turning black holes and D-branes inside out of their photon spheres,''
Phys. Rev. D \textbf{105}, no.12, 126007 (2022)
doi:10.1103/PhysRevD.105.126007
[arXiv:2110.09579 [hep-th]].
%20 citations counted in INSPIRE as of 04 May 2025

%\cite{Bianchi:2022wku}
\bibitem{Bianchi:2022wku}
M.~Bianchi and G.~Di Russo,
%``Turning rotating D-branes and black holes inside out their photon-halo,''
Phys. Rev. D \textbf{106}, no.8, 086009 (2022)
doi:10.1103/PhysRevD.106.086009
[arXiv:2203.14900 [hep-th]].
%21 citations counted in INSPIRE as of 16 May 2025

%\cite{Cipriani:2025ini}
\bibitem{Cipriani:2025ini}
A.~Cipriani, A.~De Santis, G.~Di Russo, A.~Grillo and L.~Tabarroni,
``Hamiltonian Neural Networks approach to fuzzball geodesics,''
[arXiv:2502.20881 [hep-th]].
%1 citations counted in INSPIRE as of 04 May 2025

%\cite{Bianchi:2025uis}
\bibitem{Bianchi:2025uis}
M.~Bianchi, G.~Dibitetto, J.~F.~Morales and A.~Ruip\'erez,
``Rotating Topological Stars,''
[arXiv:2504.12235 [hep-th]].
%2 citations counted in INSPIRE as of 12 May 2025

%\cite{Misner:1973prb}
\bibitem{Misner:1973prb}
C.~W.~Misner, K.~S.~Thorne and J.~A.~Wheeler,
``Gravitation,''
W. H. Freeman, 1973,
ISBN 978-0-7167-0344-0, 978-0-691-17779-3
%667 citations counted in INSPIRE as of 16 Dec 2024

%\cite{Bini:2006ime}
\bibitem{Bini:2006ime}
D.~Bini, F.~de Felice and A.~Geralico,
``Strains in General Relativity,''
Class. Quant. Grav. \textbf{23}, 7603-7626 (2006)
doi:10.1088/0264-9381/23/24/028
[arXiv:1408.4283 [gr-qc]].
%10 citations counted in INSPIRE as of 13 May 2025


%\cite{Bini:2007gxn}
\bibitem{Bini:2007gxn}
D.~Bini, F.~de Felice and A.~Geralico,
``Strains and axial outflows in the field of a rotating black hole,''
Phys. Rev. D \textbf{76}, 047502 (2007)
doi:10.1103/PhysRevD.76.047502
[arXiv:1408.4592 [gr-qc]].
%11 citations counted in INSPIRE as of 13 May 2025

%\cite{Bini:2007zzf}
\bibitem{Bini:2007zzf}
D.~Bini, F.~de Felice and A.~Geralico,
``Strains in relativity: Flat space-time analysis,''
Nuovo Cim. B \textbf{122}, 225-229 (2007)
doi:10.1393/ncb/i2007-10363-1
%0 citations counted in INSPIRE as of 13 May 2025

%\cite{Bini:2007hd}
\bibitem{Bini:2007hd}
D.~Bini, F.~de Felice and A.~Geralico,
``Strains and jets in black hole fields,''
EAS Publ. Ser. \textbf{30}, 111-117 (2008)
doi:10.1051/eas:0830011
[arXiv:0712.2396 [gr-qc]].
%0 citations counted in INSPIRE as of 13 May 2025

%\cite{Bini:2008zzb}
\bibitem{Bini:2008zzb}
D.~Bini, F.~de Felice and A.~Geralico,
``Strains in general relativity: Applications to Kerr spacetime,''
AIP Conf. Proc. \textbf{966}, no.1, 235-240 (2008)
doi:10.1063/1.2837001
%0 citations counted in INSPIRE as of 13 May 2025


\bibitem{math37} 
%\cite{Mathisson:1937zz}
%\bibitem{Mathisson:1937zz}
M.~Mathisson,
``Neue mechanik materieller systemes,''
Acta Phys. Polon. \textbf{6}, 163-200 (1937)
%549 citations counted in INSPIRE as of 13 May 2025

\bibitem{papa51} 
A. Papapetrou, 
``Spinning Test-Particles in General Relativity. I,''
{Proc.\ R.\ Soc.\ A} {\bf 209}, 248-258 (1951).
doi: 10.1098/rspa.1951.0200



\bibitem{tulc59} 
W. Tulczyjew, 
``Motion of Multipole Particles in General Relativity Theory,''
{Acta\ Phys.\ Polon.} {\bf 18}, 393 (1959).

\bibitem{dixon64} 
%\cite{Dixon:1964cjb}
%\bibitem{Dixon:1964cjb}
W.~G.~Dixon,
``A covariant multipole formalism for extended test bodies in general relativity,''
Nuovo Cim. \textbf{34}, no.2, 317-339 (1964)
doi:10.1007/BF02734579
%284 citations counted in INSPIRE as of 13 May 2025

\bibitem{dixon69}
%\cite{Dixon:1970zza}
%\bibitem{Dixon:1970zza}
W.~G.~Dixon,
``Dynamics of extended bodies in general relativity. I. Momentum and angular momentum,''
Proc. Roy. Soc. Lond. A \textbf{314}, 499-527 (1970)
doi:10.1098/rspa.1970.0020
%457 citations counted in INSPIRE as of 13 May 2025


\bibitem{dixon70}
%\cite{Dixon:1970zz}
%\bibitem{Dixon:1970zz}
W.~G.~Dixon,
``Dynamics of extended bodies in general relativity. II. Moments of the charge-current vector,''
Proc. Roy. Soc. Lond. A \textbf{319}, 509-547 (1970)
doi:10.1098/rspa.1970.0191
%178 citations counted in INSPIRE as of 13 May 2025

\bibitem{dixon73}
W.G. Dixon, 
{Gen.\ Relativ.\ Gravit.} {\bf 4}, 199  (1973).

\bibitem{dixon74}
%\cite{Dixon:1974xoz}
%\bibitem{Dixon:1974xoz}
W.~G.~Dixon,
``Dynamics of extended bodies in general relativity III. Equations of motion,''
Phil. Trans. Roy. Soc. Lond. A \textbf{277}, no.1264, 59-119 (1974)
doi:10.1098/rsta.1974.0046
%208 citations counted in INSPIRE as of 13 May 2025

\bibitem{ehlers77} 
J. Ehlers  and E. Rudolph,
``Dynamics of extended bodies in general relativity center-of-mass description and quasirigidity,''
{Gen.\ Relativ.\ Gravit.} {\bf 8}, 197-217  (1977)
doi:10.1007/BF00763547

\bibitem{quadrup_schw}
%\cite{Bini:2013rrx}
%\bibitem{Bini:2013rrx}
D.~Bini and A.~Geralico,
``Dynamics of quadrupolar bodies in a Schwarzschild spacetime,''
Phys. Rev. D \textbf{87}, no.2, 024028 (2013)
doi:10.1103/PhysRevD.87.024028
[arXiv:1408.5261 [gr-qc]].
%19 citations counted in INSPIRE as of 13 May 2025


\bibitem{quadrup_kerr1}
%\cite{Bini:2013uwa}
%\bibitem{Bini:2013uwa}
D.~Bini and A.~Geralico,
``Deviation of quadrupolar bodies from geodesic motion in a Kerr spacetime,''
Phys. Rev. D \textbf{89}, no.4, 044013 (2014)
doi:10.1103/PhysRevD.89.044013
[arXiv:1311.7512 [gr-qc]].
%27 citations counted in INSPIRE as of 13 May 2025


\bibitem{quadrup_kerr_num}
%\cite{Bini:2014xyr}
%\bibitem{Bini:2014xyr}
D.~Bini and A.~Geralico,
``Extended bodies in a Kerr spacetime: exploring the role of a general quadrupole tensor,''
Class. Quant. Grav. \textbf{31}, 075024 (2014)
doi:10.1088/0264-9381/31/7/075024
[arXiv:1408.5484 [gr-qc]].
%14 citations counted in INSPIRE as of 13 May 2025

\bibitem{spin_dev_schw} 
%\cite{Bini:2011nhv}
%\bibitem{Bini:2011nhv}
D.~Bini, A.~Geralico and R.~T.~Jantzen,
``Spin-geodesic deviations in the Schwarzschild spacetime,''
Gen. Rel. Grav. \textbf{43}, 959 (2011)
doi:10.1007/s10714-010-1111-4
[arXiv:1408.4946 [gr-qc]].
%30 citations counted in INSPIRE as of 13 May 2025


%\cite{Bini:2000vv}
\bibitem{Bini:2000vv}
D.~Bini, G.~Gemelli and R.~Ruffini,
``Spinning test particles in general relativity: Nongeodesic motion in the Reissner-Nordstrom space-time,''
Phys. Rev. D \textbf{61}, 064013 (2000)
doi:10.1103/PhysRevD.61.064013
%21 citations counted in INSPIRE as of 13 May 2025

%\cite{Bini:2011nhv}
\bibitem{Bini:2011nhv}
D.~Bini, A.~Geralico and R.~T.~Jantzen,
``Spin-geodesic deviations in the Schwarzschild spacetime,''
Gen. Rel. Grav. \textbf{43}, 959 (2011)
doi:10.1007/s10714-010-1111-4
[arXiv:1408.4946 [gr-qc]].
%30 citations counted in INSPIRE as of 13 May 2025

%\cite{Bini:2011tvf}
\bibitem{Bini:2011tvf}
D.~Bini and A.~Geralico,
``Spin-geodesic deviations in the Kerr spacetime,''
Phys. Rev. D \textbf{84}, 104012 (2011)
doi:10.1103/PhysRevD.84.104012
[arXiv:1408.4952 [gr-qc]].
%22 citations counted in INSPIRE as of 13 May 2025

%\cite{Callan:1985ia}
\bibitem{Callan:1985ia}
C.~G.~Callan, Jr., E.~J.~Martinec, M.~J.~Perry and D.~Friedan,
``Strings in Background Fields,''
Nucl. Phys. B \textbf{262}, 593-609 (1985)
doi:10.1016/0550-3213(85)90506-1
%1926 citations counted in INSPIRE as of 04 May 2025

%\cite{Polchinski:1998rq}
\bibitem{Polchinski:1998rq}
J.~Polchinski,
``String theory. Vol. 1: An introduction to the bosonic string,''
Cambridge University Press, 2007,
ISBN 978-0-511-25227-3, 978-0-521-67227-6, 978-0-521-63303-1
doi:10.1017/CBO9780511816079
%801 citations counted in INSPIRE as of 12 May 2025

%\cite{Becker:2006dvp}
\bibitem{Becker:2006dvp}
K.~Becker, M.~Becker and J.~H.~Schwarz,
``String theory and M-theory: A modern introduction,''
Cambridge University Press, 2006,
ISBN 978-0-511-25486-4, 978-0-521-86069-7, 978-0-511-81608-6
doi:10.1017/CBO9780511816086
%302 citations counted in INSPIRE as of 12 May 2025


%\cite{Cardoso:2008bp}
\bibitem{Cardoso:2008bp}
V.~Cardoso, A.~S.~Miranda, E.~Berti, H.~Witek and V.~T.~Zanchin,
``Geodesic stability, Lyapunov exponents and quasinormal modes,''
Phys. Rev. D \textbf{79}, no.6, 064016 (2009)
doi:10.1103/PhysRevD.79.064016
[arXiv:0812.1806 [hep-th]].
%833 citations counted in INSPIRE as of 12 May 2025


%\cite{Jantzen:1992rg}
\bibitem{Jantzen:1992rg}
R.~T.~Jantzen, P.~Carini and D.~Bini,
``The Many faces of gravitoelectromagnetism,''
Annals Phys. \textbf{215}, 1-50 (1992)
doi:10.1016/0003-4916(92)90297-Y
[arXiv:gr-qc/0106043 [gr-qc]].
%203 citations counted in INSPIRE as of 28 Mar 2025


%\cite{Bini:2012msy}
\bibitem{Bini:2012msy}
D.~Bini, A.~Geralico and R.~T.~Jantzen,
``Separable geodesic action slicing in stationary spacetimes,''
Gen. Rel. Grav. \textbf{44}, 603 (2012)
doi:10.1007/s10714-011-1295-2
[arXiv:1408.5259 [gr-qc]].
%9 citations counted in INSPIRE as of 13 May 2025

%\cite{Bianchi:2021mft}
\bibitem{Bianchi:2021mft}
M.~Bianchi, D.~Consoli, A.~Grillo and J.~F.~Morales,
``More on the SW-QNM correspondence,''
JHEP \textbf{01}, 024 (2022)
doi:10.1007/JHEP01(2022)024
[arXiv:2109.09804 [hep-th]].
%69 citations counted in INSPIRE as of 17 May 2025


\end{thebibliography}
\end{document}